\documentstyle[11pt]{article}
\def\gappeq{\mathrel{\rlap {\raise.5ex\hbox{$>$}}
{\lower.5ex\hbox{$\sim$}}}}

\def\lappeq{\mathrel{\rlap{\raise.5ex\hbox{$<$}}
{\lower.5ex\hbox{$\sim$}}}}

\date{}

\begin{document}

\setcounter{page}{0}
\thispagestyle{empty}

\begin{flushright}
BNL-63102\\
CERN-TH/96-105\\
DESY 96-088\\
hep-ph/9605425
\end{flushright}
\vspace{0.2cm}

\begin{center}
{\Large
\bf
Deep-Inelastic Final States in a}\\
{\Large
\bf
Space-Time Description of Shower Development}\\
{\Large
\bf
and Hadronization
}
\end{center}
\bigskip
\smallskip

\begin{center}

{\large
{\bf John Ellis{\boldmath $^a$}, Klaus Geiger{\boldmath $^b$}
and H. Kowalski{\boldmath $^c$}}}
\bigskip

$^a${\it Theoretical Physics Division, CERN,
CH-1211 Geneva 23, Switzerland: John.Ellis@cern.ch}

$^b${\it Physics Department, BNL, Upton, N.Y. 11973, U.S.A.: klaus@bnl.gov}

$^c${\it DESY, Notkestrasse 85, 22607 Hamburg, Germany: kowalski@desy.de}
\end{center}
\vspace{0.5cm}

\begin{center}
{\large {\bf Abstract}}
\end{center}
\smallskip
We extend a quantum kinetic approach to the description of hadronic
showers in space, time and momentum space to deep-inelastic $ep$
collisions,
with particular reference to experiments at HERA. We follow the history
of
hard scattering events back to the initial hadronic state and forward to
the
formation of colour-singlet pre-hadronic clusters and their decays into
hadrons. The time evolution of the space-like initial-state shower and
the
time-like secondary partons are treated similarly, and cluster formation
is
treated using a spatial criterion motivated by confinement and a
non-perturbative model for hadronization. We calculate the time
evolution of
particle distributions in rapidity, transverse and longitudinal space.
We
also compare the transverse hadronic energy flow and the distribution of
observed hadronic masses with experimental data from HERA, finding
encouraging results, and discuss the background to large-rapidity-gap
events. The techniques developed in this paper may be
applied in
the future to more complicated processes such as $eA, pp, pA$ and $AA$
collisions.

\noindent

\vspace{0.5cm}


\newpage

\noindent {\bf 1. INTRODUCTION}
\bigskip

The short-distance physics of isolated hard partonic processes
in high energy particle collisions is nowadays generally well
understood within perturbative QCD, either by calculating
matrix elements with parton final states,
or by parton shower evolution based on the QCD renormalization group
equation.
On the other hand, the long-distance
dynamics of non-perturbative soft processes and of the confinement
mechanism in the process of final-state parton-hadron conversion
is presently not calculable from first principles, and therefore
requires phenomenological model building.
Nevertheless, over the past two decades, the combination
of perturbative QCD calculus and realistic modelling of the
non-perturbative physics has been developed sufficiently to provide an
impressively accurate and predictive description of a large class of
experimental observables connected with large-momentum jets
\cite{lep,ua5,cteq}.

However, with the advent of HERA
($ep$, possibly $eA$) and the Tevatron ($p\bar p$), a new regime of QCD
at {\it high parton density} is opening up, with which one is just
beginning
to come to grips. This regime will be further explorable with the
future accelerators RHIC ($AA$)  and LHC ($pp$,$pA$, $AA$).
The common novel feature of these machines is the opportunity to
study the production and evolution of a system of a large number
of partons per unit phase space
$\Delta \Omega \equiv \Delta r \Delta k$,
which provides a possible
source for new phenomena such as non-trivial statistical particle
correlations, coherence and interference effects, dissipation and
collective excitations.
Examples of the experimental manifestation of such phenomena are:
in $ep$ ($eA$) collisions, an enhanced growth of the parton
distributions
at small Bjorken $x$ \cite{smallx}, as well as the observation
of diffractive events with large rapidity gaps between target and
current fragmentation regions \cite{lrg1,lrg2};
 in $eA$, $pA$, $AA$ collisions, events with multiple parton scattering
\cite{mps,ms3}, the QCD Landau-Pomeranchuk-Migdal effect \cite{lpm2}
and jet quenching \cite{jetqu};
in $AA$ collisions, the possible formation of a high-temperature,
deconfined parton plasma \cite{qgp}.

To quantify what we mean by {\it high parton density}, consider
the hard interaction of a probing particle with a hadron or nucleus via
a momentum transfer $Q\gg\Lambda_{QCD}$. The probe can be,
e.g., a photon (in deep-inelastic scattering)
or a parton (in hadronic or nuclear collisions).  The hard interaction
probes
space-time distances $r \sim 1/Q$, thereby resolving
a density of partons in the probed  hadron (nucleon) or nucleus
which may be characterized by the number of quark and gluon quanta
with a definite value of rapidity $y \simeq \ln(1/x)$
in the transverse plane,
$
\rho_{qg}\propto R_N^{-2} d N_{qg}/dy \simeq\
 R_N^{-2} \,(A\, xf_N(x,Q^2))
$,
where $f_N(x,Q^2)$ denotes the sum of quark and gluon parton
distributions
in a nucleon,
$R_N$ the nucleon radius, and $A$  the number of nucleons.
One can distinguish three regions \cite{levin94}:
(i)
$r \ll 1$ fm, $\rho_{qg} \ll R_N^{-2}$ $-$
the short-distance, low-density
regime of perturbative QCD,
(ii)
$r \approx 1$ fm $-$
the non-perturbative QCD domain of the
complex mechanism of confinement;
and (iii)
$r \ll 1$ fm, $\rho_{qg} \, \lower3pt\hbox{$\buildrel >\over\sim$}\,
R_N^{-2}$
 $-$
the high-density regime where
a  dense of parton system is probed at short distances,
so that perturbative methods may be applied, within a statistical
approach.

There are two extreme ways to penetrate a system of partons with large
density
at short space-time distances:
one way is deep-inelastic $ep$ scattering
($A=1$) at high energy in the region of very small Bjorken $x\ll 1$.
For instance, at HERA, the extrapolation of experimental
data implies 30-40(!) gluons
in a proton at $x \simeq 10^{-4}$ \cite{hera}.
The other way is through collisions of heavy nuclei,
in which one can reach high parton densities at not so very high
energies
or small $x$, due to the large number of overlapping nucleons ($A\gg
1$).
This presumably can be achieved at RHIC ($x\approx 10^{-1}$-$10^{-2}$),
and certainly
at the LHC ($x\approx 10^{-3}$-$10^{-4}$). In particular, at the LHC
both the conditions of
small $x$ and large $A$ may be combined.
It is clear that the theoretical study of high-density QCD requires the
development of new methods by recruiting techniques from
relativistic many-body physics, the kinetic theory of transport
phenomena,
renormalization group at finite density (and finite temperature), etc..

The purpose of the present paper is to start looking at this physics
from a
space-time point of view, and to study the dynamics of
high parton densities in deep-inelastic $ep$ scattering (DIS) in the
kinematical region covered by the HERA experiments ZEUS and H1.
In the light of the detailed hadronic
measurements at HERA, which provide
information about the underlying parton and hadron dynamics,
our emphasis is the study of the interplay between perturbative
partonic processes and non-perturbative hadron formation.
We employ a kinetic space-time approach
to parton-shower evolution combined with a
statistical model of parton-hadron conversion \cite{ms37} that allows us
 to follow
the time development of the particle system in both momentum space and
position space, i.e., in 7-dimensional phase space $d^3r d^4k$.

The space-time structure
of the production and evolution of partonic colour charges and their
conversion
into `white' hadrons is the key problem in the dynamics of complex
multi-parton
systems.
In the context of high-density QCD, insight into this problem is
especially important,
because the presence of many partons close by in phase space, generated
by the
particle dynamics itself, necessarily causes the propagation and
interaction of quanta to become non-local and to be
correlated statistically in position space and colour space.
As the system evolves, these conditions will change with time and will
in general depend on the local density of particles.
This is to be contrasted with the familiar translation-invariant
evolution
of well-separated parton jets in empty space, in which case
space-time correlations are absent or irrelevant,
because the jets evolve undisturbed by each other.
An interesting  example of a deviation from
unscathed jet evolution has been conjectured to occur
in $e^+e^-$ annihilation
into hadrons via $W^+W^-$ production \cite{CERNyr}, where the jets from
the two
 $W$'s
overlap and crosstalk, so that
the interplay between space-time dynamics and the
colour flow of close-by partons may lead to a noticeable shift in the
experimental $W$ mass determination \cite{ms40}.
\smallskip

\noindent
Summarizing the above arguments, our motivation in the following
is two-fold:
\begin{description}
\item{(i)}
First, we would like to provide an alternative and supplementary
analysis of standard non-diffractive DIS events, in order to
estimate the `background'
\footnote{
The term  `background' is not to be understood literally, because
the contribution of diffractive events with a large rapidity gap
at HERA is of the order of 10 \%, which is still
comparably small, although experimentally significant \cite{lrg1}.}
to the less well understood diffractive, `large rapidity gap' events.
To the extent that other parton shower models
\cite{mc,ariadne,herwigmc,pythia,lepto,ing88} are generally describing
this background well, our approach should give the same answer,
because our additional space-time information which is not contained in
previous
investigations should not contradict the well-known parton evolution
in momentum space.
On the other hand, the space-time geography of non-diffractive events
may shed some light on the dynamics of diffractive
events, which presumably undergo a different space-time development.
\item{(ii)}
Secondly,
since our approach is in principle designed to be
universally applicable to high-energy collisions involving
lepton, hadron, or nuclear beams,  we also see $ep$ collisions at
HERA as  a learning ground for future $eA$ (HERA?), $pp$, $pA$, $AA$
(RHIC, LHC) experiments, whose theoretical description
certainly requires knowledge of space-time evolution in order
to resolve the complex multi-particle dynamics over the expected long
collision time scales.
\end{description}
\smallskip

The paper is organized as follows. In Section 2 we review specific
features of DIS at HERA,
with the primary aim of establishing our nomenclature and notation for
kinematic variables. In Section 3 we introduce the general concept of
our
model for the space-time development of the hadronic system, recalling
relevant aspects of our framework of quantum multiparticle kinetics, the
treatment of the initial state, the space-time development of the
space-like
and time-like parton showers associated with initial- and final-state
radiation, and our spatial criterion for the formation of hadronic
clusters
and their subsequent decays. Section 4 presents our main results,
including
the time development of the rapidity distribution, inclusive hadronic
spectra and transverse energy flow. Particular attention is paid to the
distribution of the mass $M_X$ of the observed hadronic final state
in events
without large rapidity gaps, which reflects the details of our cluster
formation mechanism and hadronization procedure.
\bigskip
\bigskip

\noindent {\bf 2. SPECIFIC FEATURES OF DIS AT HERA}
\bigskip

For the purpose of clarity and to define quantities used subsequently,
we briefly review in this Section some basic notions and kinematics,
 focusing on the conditions of
the $ep$ collider HERA, where
an electron beam and a proton beam with
four momenta $p_{e,\,p} \equiv (E, p_z, \vec 0_\perp)_{e,\,p}$ and
\begin{equation}
E_e \;=\; 27 \;\mbox{GeV}
\;,\;\;\;\;\;\;\;\;
E_p \;=\; 820 \;\mbox{GeV}
\;,\;\;\;\;\;\;\;\;
\sqrt{s}\;=\;296 \;\mbox{GeV}
\label{EEs}
\end{equation}
collide head-on.
For comparison, in the centre of mass of electron and proton,
the energies are $E_e\simeq E_p = 148$ GeV, corresponding to
a global shift of the proton rapidity as compared to (\ref{EEs})
from $\vert y_p \vert= 7.45$ to $\vert y_p\vert = 4.45$.
\medskip

\noindent {\bf 2.1 Event types}
\smallskip

\noindent
The physics at HERA may be separated in two classes of event types,
illustrated in Fig. 1,
whose definitions are:
\begin{description}
\item{(i)}
{\it Non-diffractive events} (Fig. 1a): Here the exchanged virtual
photon
\footnote{In the kinematic region investigated, contributions from
$Z^0$ exchange can be neglected.}
destroys the coherence of the incoming proton by a hard scattering off a
quark inside the proton, and breaks up the proton into coloured
subsystems,
which are essentially a jet (led by the struck quark) and the proton
remnant system
(consisting of the partons that have not taken part in the hard
interaction).
This class of events is well described by the standard QCD hard
scattering
picture.
\item{(ii)}
{\it Diffractive events} (Fig. 1b): This class is characterized by
an interaction in which the proton either remains intact or
receives some small internal excitation to become a relatively
low-mass system, and in which the virtual photon also fragments
into a relatively low-mass system of particles.
This leads generally to experimentally-observable large rapidity gaps
\footnote{
The terms `diffractive events' and `large rapidity-gap events' are
often used synonymously.}
between the outgoing proton and the rest of the produced hadronic
system,
which may be interpreted as the exchange of a colourless object
(the `pomeron') between the photon and the proton.
\end{description}

For the remainder of this paper we consider exclusively
the non-diffractive event type, which is describable from first
principles in
terms of the perturbative QCD parton picture, and for which
our space-time approach in terms of photon-quark
hard scattering, parton shower evolution and parton-hadron conversion
is applicable as an extension of our previous work on
$e^+e^-$-collisions \cite{ms39,ms40}.
The diffractive event type will not be addressed here, since
it requires specific model extensions which we want to avoid at this
point.
\medskip

\noindent {\bf 2.2 Kinematics}
\smallskip

\noindent
The pecularities of the kinematics of DIS in general, and of the HERA
facility
in particular, require a clear specification of which Lorentz frame
is chosen - an issue which is especially important when dealing
with the space-time dynamics.
The HERA {\it laboratory frame} ($\equiv$ $ep$ $lab$) is the actual
experimental setup (c.f. Fig. 2a),
in which electron and proton beams collide head on, but
with beam momenta that differ by more than an order of magnitude.
This is different from the $ep$ {\it centre-of-mass frame}
($\equiv$ $ep$ $cms$)
in which electron and proton have equal but opposite  momentum,
and which is shifted in rapidity as compared to the laboratory frame.
Most convenient for theoretical analyses, however, is
the
$\gamma p$ {\it centre-of-mass frame} ($\equiv$ $\gamma p$ $cms$), in
which
the virtual photon and proton collide head on (c.f. Fig. 2b).

Our convention  in the following is that
frame-dependent quantities generally refer to the $ep$ $lab$ or the
$ep$ $cms$ (which, as mentioned after (\ref{EEs}), is related to the
former by a trivial shift of the
proton rapidity by 2.3 units), whereas
Lorentz non-invariant quantities which refer to the
$\gamma p$ $cms$ are marked by an asterisk.
For instance,
$E$ and $k_\perp$
represent a particle's energy and the momentum
transverse to the electron-proton axis in the $ep$ system, respectively,
while $E^\ast$ denotes the energy in the $\gamma p$ $cms$ and
$k_\perp^\ast$
the momentum transverse to the photon-proton axis.
In either frame, we define the negative $z$-axis by the proton
direction.

Let $p_e$ ($P$) denote the electron (proton)
incoming momenta
and  $q$ the space-like photon 4-momentum,
and define the standard Lorentz invariants for DIS
 as
\begin{equation}
s \;\equiv\; (p_e\, +\, P)^2
\;,\;\;\;\;\;\;\;\;
Q^2 \;\equiv\; - q^2
\;,\;\;\;\;\;\;\;\;
x\;\equiv\;\frac{Q^2}{2 P\cdot q}
\;,\;\;\;\;\;\;\;\;
y\;\equiv\;\frac{P \cdot q}{P\cdot p_e}
\;,
\label{sxy}
\end{equation}
in terms of these measured momenta,
where $s$ is the total invariant mass squared of the $ep$ system
(\ref{EEs}),
$Q^2$ specifies the invariant mass of the photon,
and $x$, $y$ are the usual dimensionless Bjorken variables,
commonly termed the `scaling variable' and the `inelasticity parameter',
respectively.
>From these definitions, one finds for $Q^2 \gg M_p^2$ that
\begin{equation}
Q^2 \;\approx\; x\,y\, s
\;,\;\;\;\;\;\;\;\;
W^2\;\equiv\; (P\, +\, q)^2 \; \approx \,Q^2 \, \frac{1 - x}{x}
\;,
\label{QW}
\end{equation}
where $W$ is the invariant mass of the  hadronic
system, which equals the total $cm$ energy in the $\gamma p$ $cms$.
Table 1 familiarizes the kinematic relations
among variables $x$, $Q^2$, $y$ and $W^2$
with some numerical examples.
Fig. 3 presents schematically the phase-space regime spanned
by these variables, and emphasizes the region in
the $x$-$Q^2$ plane which is experimentally investigated at HERA.
The region of former fixed-target experiments is also indicated,
corresponding to $y < 0.01$ and $Q^2 \lappeq 100\; GeV^2$.

For the purpose of relating the kinematic conditions in the
$ep$ $lab$ to the experimental observables measured or calculated
in the $\gamma p$ $cms$, we need the Lorentz transformation
of the particle 4-vectors $p_\mu$ and $p_\mu^\ast$. For instance,
the four-momenta  of the incoming proton and photon, and of the incoming
and
outgoing (struck) quark, respectively,
are in the $ep$ $lab$ (Fig. 2a) given by
\begin{eqnarray}
P &=& \left(E_p, \;0, \;0, \,- E_p \right)
\nonumber \\
q &=& \left( y E_e - \frac{Q^2}{4E_e}, -
\sqrt{(1-y) Q^2}, \; 0, -yE_e - \frac{Q^2}{4E_e} \right)
\nonumber \\
p_q &=& \left(x\,E_p,\;0,\;0,\;x\,E_p \right)
\nonumber \\
k_q &=& \left( y E_e + \frac{(1-y)Q^2}{4y E_e}, -
\sqrt{(1-y) Q^2}, \; 0,
-yE_e + \frac{(1-y)Q^2}{4yE_e} \right)
\;,
\label{labframe}
\end{eqnarray}
where $Q \gg M_p$ is assumed.
On the other hand, in the preferable $\gamma p$ $cms$ (Fig. 2b),
the corresponding momenta are
\begin{eqnarray}
P^\ast &=& \frac{2 y E_e E_p }{\sqrt{4yE_eE_p
-Q^2}}\,\left(1,\; 0,\;0,\;1\right)
\nonumber \\
q^\ast &=& \frac{2 y E_e E_p }{\sqrt{4yE_eE_p
-Q^2}} \,
\left(1 - \frac{Q^2}{2yE_eE_p}, \;0,\;0, \; - 1\right)
\nonumber \\
p_q^\ast &=& x\;P^\ast
\nonumber \\
k_q^\ast &=& \frac{2 y E_e E_p }{\sqrt{4yE_eE_p
-Q^2}} \,
\left(1 - \frac{Q^2}{4yE_eE_p}, \;0,\;0,
\,-1 + \frac{Q^2}{4yE_eE_p} \right)
\;.
\label{hcmframe}
\end{eqnarray}

The invariant differential  cross-section for non-diffractive events
with a hard photon scattering is the convolution of the elementary
photon-quark cross-section with the quark and antiquark densities in the
struck proton,
\begin{equation}
\frac{d \sigma}{dx dy}
\;=\;
\;\,\sum_{i}
\frac{d \hat \sigma_i}{dx dy}
\;\, e_i^2 \,x f_i(x,Q^2)
\;,
\label{xsec1}
\end{equation}
where the index $i$  labels the quark and antiquark flavors,
with $e_i$ and $f_i(x,Q^2)$ denoting the
corresponding electric charges and (anti)quark distributions of the
proton.
The associated elementary cross-sections $d\hat \sigma_i$
are given to lowest order (i.e., before any QCD radiation) by
\begin{eqnarray}
\frac{d \hat \sigma_i}{dx dy}
&=&
\frac{2\pi \alpha_{em}^2 \,e_i^2}{y Q^2}
\;\left\{
\left(1 + (1-y)^2\right)\;+\; \frac{4 p_{\perp\,prim}^2}{Q^2}\,(1-y)
\right.
\nonumber \\
& &
\;\;\;\;\;\;\;\;\;\;\;\;\;\;\;\;
\left.
\;-\;\frac{4 p_{\perp\,prim}}{Q}\,(2-y)\,\cos (\phi)
\;+\;\frac{4 p_{\perp\,prim}^2}{Q^2}\,(1-y)\,\cos (2\phi)
\right\}
\;,
\label{xsec2}
\end{eqnarray}
where $\vec{p}_{\perp\,prim} = p_{\perp\,prim} (\cos\phi, \sin\phi)$
is the intrinsic transverse momentum of the primary, initial
quark or antiquark due to the Fermi motion of the partons inside the
proton.
Intuitively, one expects the value of $p_{\perp\,prim}$ to be of the
order of
the inverse proton radius, and it is in fact
determined experimentally in hadronic collisions as well as in DIS to be
$\approx$ 400-450 MeV \cite{pTprim}.
\bigskip
\bigskip
\newpage

\noindent {\bf 3.  THE MODEL}
\bigskip

\noindent {\bf 3.1  General concept}
\smallskip

The central element in our approach is the use of
QCD transport theory \cite{msrep} and quantum field kinetics \cite{ms39}
to follow the evolution of a generally mixed multiparticle system
of partons and hadrons
in 7-dimensional phase-space $d^3 r d^3 k dk^0$.
We include both the perturbative QCD parton-cascade development
\cite{MLLA1,MLLA2,jetcalc,bassetto},
and the phenomenological
parton-hadron conversion model which we have proposed previously in
Refs.
\cite{ms37,ms40},
in which we consider dynamical parton-cluster formation as
a local, statistical process that depends on the spatial separation and
colour
of nearest-neighbour partons, followed by the decay of clusters into
hadrons.
In contrast to the commonly-used momentum-space description,
in our approach we trace
the microscopic history of the dynamically-evolving particle system
in space-time {\it and} momentum space, so that
the correlations of partons in space,  time, and colour can be taken
into account for both the perturbative cascade evolution
and the non-perturbative hadronization.
We emphasize that one strength of this approach lies in the possible
extension of
its
applicability to the collision dynamics of
complicated multi-particle systems, as in $eA$, $pA$
and $AA$ collisions, for which a causal time evolution in position space
and momentum space is essential.

The model contains
three main building blocks which generically embody
high-energy collisions involving leptons, hadrons, or nuclei in
colliders
(for DIS $ep$ collisions, the model components are illustrated in Fig.
4):
\begin{description}
\item[a.]
the {\it initial state} associated with the
incoming collision partners (the beam particles),
in particular the phenomenological construction of the hadron (nucleus)
in terms of quark and gluon phase-space distributions;
\item[b.]
the {\it parton cascade development}
with mutual- and self-interactions of the system of quarks and gluons
consisting of both the materialized partons from parton showers,
and the spectator partons belonging to the remnants of the
collided beam particles;
\item[c.]
the {\it hadronization} of the evolving system
in terms of parton coalescence to colour-neutral clusters
as a local, statistical process that depends on the spatial separation
and colour
of nearest-neighbour partons, followed by the decay of clusters into
hadrons
according to the density of final hadron states.
\end{description}
Such a pragmatical division, which assumes complex interference
between the different physics regimes to be negligible, is possible if
the respective dynamical scales are such that
the short-range hard interaction, with its associated perturbative
parton evolution, and the non-perturbative
mechanism of hadron formation occur on well-separated  space-time
scales.
For DIS, this condition of validity requires
$\min ( W^2, Q^2 )\ge L_c^{-2} \gg \Lambda^2_{QCD}$,
meaning that the characteristic mass scale for
the $\gamma p$ hard scattering and parton shower development
($W^2$, $Q^2$, or a combination of the two) is larger than
the inverse `confinement length scale' $L_c = O(1\,fm)$
separating perturbative and non-perturbative domains.
Specifically, for DIS, it is apparent from
(\ref{QW}) that in the small-$x$ regime
probed at HERA ($10^{-4} \, \lower3pt\hbox{$\buildrel <\over\sim$}\,
x \, \lower3pt\hbox{$\buildrel <\over\sim$}\, 10^{-3}$),
one has
$60 \, \lower3pt\hbox{$\buildrel <\over\sim$}\,
W \, \lower3pt\hbox{$\buildrel <\over\sim$}\, 300$ GeV for
$10 \le Q^2 \le 300$ GeV$^2$, so that the above requirement is well
satisfied.
We emphasize however, that in our model the interplay
between perturbative and non-perturbative regimes is controlled locally
by the space-time evolution of the mixed parton-hadron system itself,
rather than by an arbitrary global division
between parton and hadron degrees of freedom.
\smallskip

We now turn to the specific case of DIS, and in the following
subsections we will discuss the above components in more detail.
\medskip

\noindent {\bf 3.2  Framework of quantum kinetics for multiparticle
dynamics}
\smallskip

>From quantum kinetic theory, one can obtain
a space-time description of multiparticle systems
in high-energy QCD processes, as has been discussed formally
in Ref. \cite{ms39}.
Applied to the concept of our model, as outlined in Sec. 3.1,
this framework allows us to
express the time evolution of the mixed system of
incoherent partons, composite clusters, and physical hadrons
in terms of a closed set of
integro-differential equations for
the local phase-space densities of the different particle excitations.
The definition of these phase-space densities (`Wigner densities'),
denoted by
$F_\alpha$, where $\alpha\equiv p, c, h$
labels the species of partons, prehadronic clusters, or hadrons,
respectively, is:
\begin{equation}
F_\alpha(r,k)\;\,\equiv\; \, F_\alpha (t, \vec r; E, \vec k)
\;\,=\;\,
\frac{dN_\alpha (t)}{d^3r d^3k dE}
\;,
\label{F}
\end{equation}
where $k^2 = E^2 -\vec{k}^{\,2}$ can be off or on mass shell.
The densities (\ref{F}) measure the number of particles
of type $\alpha$ at time $t$ with position in $\vec r + d\vec{r}$,
momentum in $\vec k + d\vec{k}$,
and energy in $E + dE$ (or equivalently invariant mass in $k^2 + dk^2$).
The $F_\alpha$ are the quantum analogues of the
classical phase-space distributions, and
contain the essential microscopic
information required for a statistical description
of the time evolution of a many-particle system in
complete phase space, thereby providing the basis for calculating
macroscopic observables
in the framework of relativistic kinetic theory.

The Wigner densities (\ref{F}) are determined by the
self-consistent solutions of
a set of transport equations (in space-time) coupled with
renormalization-group-type equations (in momentum space).
Referring  to Refs. \cite{ms40,ms37,ms39} for details,
we remark that
these equations can be generically expressed as
convolutions of the densities of radiating or interacting particles
$F_\beta$
with  specific cross sections $\hat{I}_j$ for the processes $j$,
yielding the following closed set of balance equations for the
space-time development of
the densities of partons $F_{p}$, clusters $F_c$ and
hadrons $F_h$,
\begin{eqnarray}
k \cdot \partial_r \; F_p(r,k)
&=&
F_{p'}\circ \hat{I}(p'\rightarrow p p'')\;-\;
F_p \circ\hat{I}(p\rightarrow p' p'')  \;-\;
F_p\,F_{p'}\circ \hat{I}(p p'\rightarrow c)
\label{e1}
\\
k\cdot\partial_r \; F_c(r,k)
&=&
F_p\,F_{p'}\circ \hat{I}(p p'\rightarrow c)
\;-\;
F_c\circ \hat{I}(c\rightarrow h)
\label{e2}
\\
k\cdot \partial_r \; F_h(r,k)
&=&
F_c \circ\hat{I}(c\rightarrow h)
\label{e3}
\;,
\end{eqnarray}
where $k\cdot \partial_r \equiv  k_\mu \partial/\partial r^\mu$.
We remark that eq. (\ref{e1}) implicitly embodies the momentum space
($k^2$) evolution of partons through
the renormalization of the phase-space densities $F_p$, determined
by their change $k^2 \partial F_p(r,k)/\partial k^2$
with respect to a variation of the mass (virtuality) scale $k^2$
in the usual QCD evolution framework \cite{MLLA1,MLLA2}
\footnote{
For pre-hadronic clusters and hadrons, we assume renormalization effects
to be comparatively small, so that their
mass fluctuations $\Delta k^2/k^2$ can be ignored to first
approximation,
implying $k^2 \partial F_c(r,k) /\partial k^2
= k^2 \partial F_h(r,k) / \partial k^2  =0
$.
}.
Each of the terms on the right-hand side
of (\ref{e1})-(\ref{e3})
corresponds to one of the following categories (c.f. Fig. 4):
(i)
parton multiplication through radiative emission processes
on the perturbative level,
(ii)
colourless cluster formation through parton recombination
depending on the local colour and spatial configuration,
(iii)
hadron formation through decays of the cluster excitations
into final-state hadrons.
Each convolution $F \circ\hat{I}$ of
the density of particles $F$ entering a particular vertex
${\hat I}$ includes a sum over contributing
subprocesses, and a phase-space integration
weighted with the associated subprocess probability distribution
of the squared amplitude.

The  equations (\ref{e1})-(\ref{e3}) reflect a probabilistic
interpretation of QCD evolution in space-time and momentum space
in terms of sequentially-ordered interaction processes $j$,
in which the rate of change of the particle distributions $F_\alpha$
($\alpha=p,c,h$)
in a phase-space element $d^3rd^4k$
is governed by the balance of gain (+) and loss ($-$) terms.
The left-hand side
describes free propagation of a
quantum of species $\alpha$, whereas
on the right-hand side the interaction kernels $\hat{I}$
are integral operators that incorporate the effects of
the particles' self  and mutual interactions.
This quasi-classical, probabilistic  character of high-energy particles
is essentially an effect of time dilation, because in any frame
where the particles move close to the speed of light, the associated
wave-packets are highly localized to short space-time extent, so that
long-distance quantum interference effects are generally very small.
\medskip

\noindent {\bf 3.3  Scheme of solution and choice of Lorentz frame}
\smallskip

In the above kinetic approximation \cite{ms39} to the multi-particle
dynamics,
the probabilistic character of the evolution equations
(\ref{e1})-(\ref{e3})
allows one to solve for the Wigner densities $F_\alpha(r,k)$ by
simulating
the dynamical development as a Markovian process causally in time.
Because it is an initial-value problem, one must specify
some physically appropriate initial
condition $F_\alpha(t_0,\vec{r},k)$ at starting time $t_0$, such that
all the dynamics prior to this point is effectively embodied in this
initial form of $F_\alpha$.
The set of kinetic equations (\ref{e1})-(\ref{e3}) can
then be solved
in terms of the evolution of the Wigner densities $F_\alpha$
for $t > t_0$ using Monte Carlo methods to
simulate the time development of the mixed system
of partons, clusters, and hadrons
in position and momentum space \cite{ms40,msrep}.

In the next subsections we explain in more detail the
different components for the case of DIS,
namely, the initial-state Ansatz, parton shower
development, and parton-hadron conversion.
The overall  {\it concept of the simulation} is
illustrated in Fig. 4 and can be summarized as follows:
given the initial state of the photon
and the proton disassembled into its parton content,
the hard interaction of the photon with one of the
quarks occurs at time $t=0$.
Specifying the initial state at some earlier time $t_0 < 0$, and with
the
hard scattering variables chosen from the cross-section,
the phase-space distribution of particles at $t=0$ can be
calculated and then evolved in small time steps forward,
until stable final-state hadrons are left as freely-streaming particles.
The size of time steps is chosen as
$\Delta t =O(10^{-3}\;fm)$, so that
an optimal resolution of the particle dynamics in space and
energy-momentum is achieved.
The partons propagate along classical trajectories until they interact,
i.e., decay (branching process) or recombine (cluster formation).
Similarly, the clusters so formed
travel along classical paths until they convert into
hadrons (cluster decay).
The corresponding probabilities and time scales of interactions are
sampled stochastically from the relevant probability distributions
in the kernels $\hat{I}$ of eq. (\ref{e1})-(\ref{e3}).
\smallskip

It is clear that
the description of particle evolution is
Lorentz-frame dependent, and a suitable reference frame
must be chosen (not necessarily the laboratory frame).
When computing Lorentz-invariant quantities, such as
cross sections or final-state hadron spectra, the particular choice is
irrelevant, whereas for non-invariant observables, such as energy
distibutions
or space-time-dependent quantities, one must at the end transform
from the arbitrarily-chosen frame of theoretical description to the
actual frame of measurement.
Furthermore,  at HERA even experimental analyses are
often carried out in the $\gamma p$ $cms$ (\ref{hcmframe}),
rather than the $ep$ $lab$ (\ref{labframe}).
For our purposes it is most convenient
to choose the overall centre-of-mass frame of the colliding electron
and proton, the $ep$ $cms$, as the global frame with respect to which
the
evolution of the collision system is followed
\footnote{
However, to make contact with the HERA experiments, most of our results
will be discussed later in the $\gamma p$ $cms$ rather than the
$ep$ $cms$, unless specified otherwise.
}.
Recall our convention that the
$ep$ collision axis defines the $z$-axis, with the electron (proton)
moving in the positive
(negative) $z$ direction.  The incoming 4-momenta $p_e$ and
$P$ involve therefore no transverse components, and are
\footnote{
We emphasize that in our calculations we use  exact kinematics,
and take into account proton, electron and quark masses.
}
\begin{eqnarray}
p_e &=&
\left(\frac{s + m_e^2-M_p^2}{2\sqrt{s}},\,0,\,0,\,+ P_{cm}\right)
\;\approx\;\frac{\sqrt{s}}{2} \,\left(1,\,0,\,0,\,1\right)
\nonumber \\
P &=&
\left(\frac{s - m_e^2+M_p^2}{2\sqrt{s}},\,0,\,0,\,- P_{cm}\right)
\;\approx\;\frac{\sqrt{s}}{2} \,\left(1,\,0,\,0,\,-1\right)
\;,
\label{epcmframe}
\end{eqnarray}
where
$P_{cm}=\sqrt{s- (m_e+M_p)^2}\sqrt{s-(m_e-M_p)^2}/(2\sqrt{s})$ is
the $ep$ $cm$ momentum.
\medskip

\noindent {\bf 3.4  Initial state}
\smallskip

The incoming electron is considered as a point-like object
carrying the full beam energy, meaning
that we neglect any QED or QCD substructure of the electron,
as well as initial-state photon radiation by the electron.
We assume that the electron emits the virtual photon of invariant mass
$Q^2=-q^2$ at time $t = -Q^{-1}$, so that $t=0$ characterizes the point
when the photon hits the incoming proton, as is depicted in
Fig. 4.

The incoming proton, on the other hand, is
decomposed into its parton substructure by
phenomenological construction of the
momentum and spatial distributions of its daughter partons
on the basis of the experimentally-measured proton structure functions
and elastic proton form factor.
Here it is important to distinguish between the scales $Q^2$ and $Q_0^2$
(c.f. Fig. 4):
The hard scattering scale $Q^2 = - q^2$ is set by the momentum transfer
$q$ between electron and proton and determines the parton structure
as seen by the virtual $\gamma$ {\it after} the initial state
radiation of the struck quark.
The initial resolution scale $Q_0^2$, on the other hand, determines
how detailed the parton phase-space density in the proton
would be resolved {\it before} the initial state radiation.
Hence, in accord with (\ref{F}), we
introduce the initial parton phase-space distribution
$F_a^{(0)}(r,p)$ as the number density of partons in a phase-space
element
$d^3r d^3p dE$ at time $t=t_0$ within in the
proton at an initial resolution scale $Q_0^2 = 1$ $GeV^2$.
We assume the following  factorized form:
\begin{equation}
F_a^{(0)}(r,p)\; \equiv \;
F_a(r,p) \left.\frac{}{}\right|_{t=t_0, \;\vert p^2\vert\simeq Q_0^2}
\;=\;
P_a (\vec p,\vec P; Q_0^2) \circ R_a (\vec p, \vec r,\vec R)
\left.\frac{}{}\right|_{t=t_0}
\;.
\label{Fa0}
\end{equation}
The right-hand side $P_a\circ R_a$ is a
convolution of an initial momentum distribution $P_a$ and a spatial
distribution $R_a$,
with the subscript $a= g, q_i, \bar{q}_i$
labeling the parton species (gluons or
(anti)quarks of flavor $i=1,\ldots n_f$).
The 4-vectors
$p\equiv p^\mu=(E,\vec p)$ and $r \equiv r^\mu = (t, \vec r)$
refer to the partons, whereas $\vec P = (0,0,-P_{cm})$ and
$\vec R\equiv \vec 0$ refer to the
initial 3-momentum and the position of the parent proton at $t = t_0$
in the $ep$ $cms$.
The partons' energies
$E =  \sqrt{\vec{p}^{\,2}  - Q_0^2}$
take into account initial space-like virtualities $p^2<0$,
which  reflects  the fact that
before the collision the partons are confined inside the parent proton
and cannot be treated as free particles (meaning that they do not have
enough energy to be on mass shell, but  are space-like off-shell).
The initial momentum distribution is taken as
\begin{equation}
P_a (\vec p,\vec P; Q_0^2)\;=\;
\left(\frac{x}{\tilde x}\right) \; f_a (x,Q_0^2) \;\, g (\vec p_{\perp})
\;\,\delta\left(P_z \,-\,\frac{\sqrt{s}}{2}\right) \;
\delta^2\left(\vec P_\perp\right)
\;.
\label{Pa}
\end{equation}
Here $x$ and $\tilde x$ are the partons' longitudinal momentum and
energy fractions, respectively,
\begin{equation}
x \;=\; \frac{p_z}{P_z}
\;\;;\;\;\;\;\;\;\;\;\;
\tilde x \;=\; \frac{E}{E_p}\;=\; \sqrt{x^2 +
\frac{ p_\perp^2 - Q_0^2}{E_p^2}}
\label{xx}
\end{equation}
and
the functions $f_a (x,Q_0^2)$ are the usual (measured)
quark and gluon structure functions of the proton
\footnote{
We use the GRV structure function parametrization \cite{grv},
which describes quite accurately the HERA data even at low $Q^2$ and
very small $x$.
},
which specify the longitudinal momentum
distribution, whereas the transverse momentum distribution
$g (\vec p_{\perp}) =  (2\pi p_0^2)^{-1}
\exp \left[- \vec p_\perp^2/p_0^2\right]$
takes into account the uncertainty of the transverse momentum
(``Fermi motion") due to the fact that the initial
partons are  confined within the proton.
The latter is inferred from experimental analyses
\cite{pTprim},
with $p_0 = 0.42$ GeV, corresponding to the mean
primordial transverse parton momentum
$\langle \vert \vec{p}_\perp \vert\rangle$.
The normalization is such that
\begin{eqnarray}
& &
\sum_a \int_0^1 dx \; x f_a (x, Q_0^2) \;=\; 1
\;\;\;\;\;\;\;\;\;
\int_0^\infty d^2 p_\perp \;g(\vec p_\perp)  \;=\; 1
\label{norm1}
\\
& &
\sum_a \int \, d p^2  \,\frac{d^3 p}{(2\pi)^3 (2 E)}
\; E \; P_a (\vec p, \vec P; Q_0^2) \;\equiv\; n (P,Q_0^2)
\label{norm2}
\;,
\end{eqnarray}
where $n(P,Q_0^2)$ has dimension 1/volume and gives the total number
density of partons in the proton with momentum $P$, when resolved
at the scale $Q_0^2$.
Finally, we impose the constraint
that the total invariant mass of the partons
equals the proton mass $M_p$,
\begin{equation}
\left( \sum_j E_j \right)^2 - \left( \sum_j p_{x_j}\right) - \left(
\sum_j p_{y_j}\right)^2 - \left( \sum_j p_{z_j}\right)^2\; = \; M_p^2
\label{invmass}
\;,
\end{equation}
where the summation $j = 1, \ldots n(P,Q^2)$
runs over all partons resolved at $Q^2$, as constrained by
(\ref{norm1}) and (\ref{norm2}).
With the partons' 3-momenta determined from the distributions
in $x$ and $\vec{p}_\perp$,
the requirement (\ref{invmass}) fixes
the relation between energy and momentum by assigning
to each parton an initial space-like virtuality such that
$p^2 = E^2 - \vec{p}^2 < 0$.
With this prescription,
the resulting distribution in $p^2$ is approximately Gaussian
with a mean value of $\sqrt{\langle p^2 \rangle} \approx 500$ MeV,
i.e. the typical initial virtuality of the partons is about $Q_0/2$.
\medskip

\noindent {\bf 3.5  Parton cascade development}
\smallskip

With the above construction of the initial state in terms of the
incoming electron and photon,
and the parton cloud of the proton, the dynamical development of the
system
can now be traced according to the kinetic equations
(\ref{e1})-(\ref{e3}),
starting from $t=0$.
In our statistical picture, the initial-state parton ensemble represents
a particular fluctuation of the proton wave function that
has developed between $t = t_0 \simeq Q_0^{-1} < 0$
and time $t=0$,
at which the photon with resolution $Q^2$ picks according to the
cross-section (\ref{xsec1}) a quark with specific flavour and
momentum $p = x P$ out of the incoming parton cloud,
while the other partons are
viewed as unaffected by the short-range $\gamma q$ interaction.
Consequently, as illustrated in Fig. 4,
the early stage of the time evolution is characterized
by two different physics elements:
a)
the parton showers initiated by the quark that is struck out
of the original proton wave function through the momentum
transfer from the virtual photon, and b) the propagation
of the remnant system consisting of the other
initial partons, that remain spectators of the hard process
and form the coherent remnant of the original proton.

For the parton shower development
we employ the well-established jet calculus \cite{jetcalc,bassetto}
based
on the `modified leading logarithmic approximation' (MLLA)
to the QCD evolution of hard processes \cite{MLLA1,MLLA2}.
A parton shower then reduces to a strictly-ordered sequence
of elementary branchings $q\rightarrow qg$, $g\rightarrow gg$,
$g\rightarrow q\bar q$, which can be described stochastically as a
Markov cascade in position and momentum space.
We distinguish  initial-state, {\it space-like} branchings
of the selected quark before it reaches the $\gamma q$ vertex, and
final-state, {\it time-like} radiation off the struck
quark after the hard $\gamma q$ interaction
\cite{ing88,bengt88}
\footnote{
This separation implies the neglect of interference between the
initial- and final-state showers, a common conceptual defect that is
approximately cured by matching on to the lowest-order
$O(\alpha_s)$ matrix element.}.
The separation into two `hemispheres' divided by the $\gamma q$ vertex
is illustated in Fig. 5:
it refers to both the chronological order
along the real time axis and to the order of emission vertices
in momentum space.
The initial quark that is picked out by the photon evolves
from the remote past $t = t_0 < 0$
towards the hard interaction by sequential branchings
$p_j \rightarrow p_{j+1}+p_{j+1}^\prime$, in each of which
one of the daughters continues with increasing space-like virtuality
$\vert p_{j+1}^2\vert >\vert p_j^{2}\vert$ (where $p_j^2, p_{j+1}^2
<0$),
while the other one acquires a time-like virtuality
$p_{j+1}^{\prime\,2} > 0$\ and may develop a time-like shower of its
own.
The space-like shower is consequently characterized by increasing
virtualities $\vert p_j^2\vert < \vert p_{j+1}^2 \vert$ ,
decreasing energies and increasing
opening angles, as the quark approaches the hard vertex at
$t=0$ with $\vert p_n^2 \vert \approx Q^2$.
Once the evolved quark has been struck by the photon,
the momentum transfer provides the outgoing quark with enough
energy-momentum to become a real excitation at $t=0$ and to obtain a
time-like virtuality $k_m^2 \approx Q^2$.
This materialized  quark initiates now a shower of sequential time-like
branchings $k_m\rightarrow k_{m-1}+k_{m-1}^\prime$
in which both daughters are time-like (i.e. $k_{m-1}^2,
k_{m-1}^{\prime\,2} >0$)
with decreasing virtualities $k_{m-1}^2 < k_m^2$, decreasing energies
and decreasing opening angles.
The branching chain continues into the remote future until
it is terminated by the hadronization, which we model as
the coalescence of neighboring partons in a cascade, followed by
conversion to hadrons (Sec. 3.7 below).

The specific feature of our approach is that, in addition to the
definite
virtuality and momentum,  each elementary vertex has a certain space and
time
position which is obtained by assuming that the partons in the shower
propagate on
straight-line trajectories in between the branchings.
In the MLLA framework, the
basic properties of both space-like and time-like showers are
determined by the DGLAP equations \cite{DGLAP}, but with essential
differences in time ordering, kinematics and the treatment of
infrared singularities associated with soft gluon emission.
\smallskip

\noindent {\bf 3.5.1 Space-like parton shower}

\noindent
As mentioned above and depicted in Fig. 5, the space-like
cascade starts at some time  $t= t_0 \simeq - Q_0^{-1}$
before the actual hard scattering at $t=0$,
with the initiating parton of virtuality
$\vert p_0^2 \vert \simeq Q_0^2 = M_p^2$  embodied in
the parton cloud of the incoming proton, and proceeds
up to $p^2 \equiv p_n^2 \simeq -Q^2$ at the hard vertex
set by the space-like photon virtuality.
The emitted partons  on the side branches, on the other hand,
are not connected directly with the $\gamma q$ vertex, but
evolve independently as time-like quanta.
In the cascade sequence both collinear and soft coherent branchings
are properly included  \cite{webber88}, if the development
of the chain is described in terms of `angular-ordering' variables
(rather than the virtualities $p_j^2$),
\begin{equation}
\tilde{p}_j^2 \;\equiv E_j^2\;\zeta_{j+1}
\;\;,\;\;\;\;\;\;\;\;\;\;\;
\zeta_{j+1} \;=\;\frac{p_0\cdot p_{j+1}^\prime}{E_0 \;E_{j+1}^\prime}
\;\simeq \;
1 - \cos \theta_{0,\;j+1}
\;\;\;\;\;\;\;\;\;(0 \le j \le n)
\;,
\label{tildep}
\end{equation}
where $p_j=(E_j,\vec{p}_j)$ and
$p_{j+1}^\prime=(E_{j+1}^\prime,\vec p_{j+1}^{\,\prime})$ are
assigned as in Fig. 5 for the $j^{th}$ branching
$p_j\rightarrow p_{j+1} p^\prime_{j+1}$.
The space-like cascade is then strictly ordered in the variable
$\tilde{p}_{j+1}^2 > \tilde{p}_j^2$, which is equivalent to the
ordering of emission angles,
$E_j \theta_{0, \;j+1^\prime} < E_{j+1} \theta_{0, \;j+2^\prime}$.

Because the presence of the external hard interaction at $t=0$ and $Q^2$
sets
a physical boundary condition on the kinematical evolution of the
cascade,
it is  technically advantageous to reconstruct the cascade backwards in
time starting from $t=0$ at the hard vertex $Q^2$ and
trace the history of the struck quark back to $Q_0^2$ at $t=t_0$.
The method used here is a space-time generalization of the
`backward evolution scheme'  \cite{bengt88,backevol}.
To sketch the procedure, consider the space-like branching
$p_{n-1} \rightarrow p_n p_n^\prime$ which is closest to the $\gamma q$
vertex in Fig. 5.
The virtualities satisfy \cite{bassetto}
$\vert p_n^2 \vert > \vert p_{n-1}^2 \vert$, and $p_{n}^2, p_{n-1}^2 <
0$
(space-like) but $p_n^{\prime \,2} > 0$ (time-like).
The relative probability for this branching to occur  between
$\tilde{p}^2$
and $\tilde{p}^2 + d\tilde{p}^2$ is given by
\begin{eqnarray}
d {\cal P}_{n-1,\,n}^{(S)} ( x_{n-1}, x_{n}, \tilde{p}^2;\,\Delta t)
&=&
\frac{d \tilde{p}^2}{\tilde{p}^2}\, \frac{d z}{z}
\,
\frac{\alpha_s\left((1-z) \tilde{p}^2\right)}{2 \pi} \,
\,
\gamma_{n\mbox{-}1 \rightarrow n n^\prime} (z)
\nonumber \\
& & \;\;\;\;\;\;\;
\times \;
\left(
\frac{F(r_{n-1}; x_{n-1}, \tilde{p}^2)}
{F(r_n; x_n, \tilde{p}^2)}
\right)
\;\,{\cal T}^{(S)}(\Delta t)
\;,
\label{PS}
\end{eqnarray}
where $x_j = (p_j)_z / P_z$  ($j=n, n-1$) are the
fractions of longitudinal proton momentum $P_z$, with
$F(r_j;x_j, \tilde{p}^2)\equiv F(r_j,p_j)$
the corresponding parton distriibutions
introduced before,
and the variables
\begin{equation}
z \;=\;\frac{E_n}{E_{n-1}} \;\simeq \;\frac{x_n}{x_{n-1}}
\;\;\;,\;\;\;\;\;\;\;\;
1\;-\;z \;=\;\frac{E^\prime_n}{E_{n-1}} \;\simeq \;
\frac{x_{n-1}-x_{n}}{x_{n-1}}
\label{tildez}
\end{equation}
specify the fractional energy or longitudinal
momentum of parton $n$ and $n^\prime$, respectively, taken away from
$n-1$.
The function $\alpha_s/(2\pi \tilde{p}^2)\, \gamma (z)$
is the usual DGLAP branching probability  in the MLLA,
with $\gamma (z)$ giving the energy distribution in the variable $z$.
The last factor in (\ref{PS}) determines the time interval in the $ep$
$cms$,
$\Delta t = t_n - t_{n-1}$, that is associated with the branching
process
$n-1\rightarrow n n^\prime$.
We take here simply
\begin{equation}
{\cal T}^{(S)}(\Delta t)
\;=\; \delta\left( \frac{x_n - x_{n-1}}{\vert p_n^2\vert}\, P_z
\;-\; \Delta t\right)
\label{delts}
\;,
\end{equation}
which accounts for the formation time of $n$ by its mother $n-1$ on
the basis of the uncertainty principle:
$\Delta t = \Delta E/|p_n^2|$, $\Delta E \simeq  (x_n - x_{n-1})\,P_z$.

The ``backwards evolution" of the space-like branching
$p_{n-1} \rightarrow p_n + p^\prime_n$ is expressed in terms of the
probability that parton $(n-1)$ did {\it not} branch between the
lower bound $\tilde{p}_0^2$, given by the initial resolution scale
$Q_0^2$, and
$\tilde{p}^2$.
In that case, parton $n$ can {\it not} originate from this branching,
but must have been produced otherwise or already been present in
the initial parton distributions.
This non-branching probability is given by the
{\it Sudakov form-factor for space-like branchings}:
\begin{equation}
S_n ( x_{n}, \tilde{p}^2, \tilde{p}_0^2;\,\Delta t)
\;=\;
\exp
\left\{
\,-\, \sum_{a}
\,
 \int_{\tilde{p}_0^2}^{\tilde{p}^2}
\,
\int_{z_- (\tilde{p}^{\prime})}^{z_+ (\tilde{p}^{\prime})}
\,
d {\cal P}_{n,\,n-1}^{(S)} ( x_n,  z, \tilde{p}^{\prime 2}; \,\Delta t)
\right\}
\;,
\label{slff2}
\end{equation}
where the sum runs over the possible species $a = g, q, \bar q$
of parton $n-1$.
The upper limit of the $\tilde{p}^2$-integration is set by
$\tilde{p}^2  \, \lower3pt\hbox{$\buildrel <\over\sim$}\,Q^2$,
associated with the scattering vertex of quark $n$ with the photon
in Fig. 5.
The limits $z_\pm$ are determined by kinematics \cite{webber88}:
$z_-(\tilde{p})= Q_0/\tilde{p}$ and
$z_+(\tilde{p})= 1 - Q_0/\tilde{p}$.
The knowledge of $S_n(x_{n}, \tilde{p}^2, Q_0^2)$ is enough to trace
the evolution of the branching closest to the hard vertex backwards
from $p_n^2$ at $t=t_n \equiv 0$ to
$p_{n-1}^2$ at $t_{n-1}= - x_n/|p_n^2|\,P_z$.
The next preceding branchings  $p_{n-2}\rightarrow p_{n-1}
p_{n-1}^\prime$,
etc., are then reconstructed in exactly the same manner
with the replacements $t_n \rightarrow t_{n-1}$,
$x_n \rightarrow x_{n-1}$, $p_n^2 \rightarrow p_{n-1}^2$, and so forth,
until the initial point  $p_0^2$ at $t_0 = -Q_0^{-1}$ is reached.
\smallskip

\noindent {\bf 3.5.2 Time-like parton shower}

\noindent
Time-like  cascades are initiated by the secondary partons, i.e., those
that  emerge from the side branches of the
initial-state radiation from the scattering quark before $t=0$,
as well as those that are produced by  final-state emission
from the scattered quark after $t=0$.
Consider the time-like cascade in Fig. 5, that
is initiated by the outgoing quark of momentum $k\equiv k_m$
emerging from the hard vertex at $t=0$
and off shell by an amount
$k^2 \equiv k_m^2 \, \lower3pt\hbox{$\buildrel <\over\sim$}\, Q^2$.

Again an angular-ordered (rather than virtuality-ordererd) time
evolution
of the cascade is employed to incorporate interference effects of soft
gluons emitted along the tree in Fig. 5.
In contrast to (\ref{tildep}), the time-like version of the angular
evolution variable  is \cite{webber86}
\begin{equation}
\tilde{k}_j^2 \;\equiv E_j^2\;\xi_{j-1}
\;\;,\;\;\;\;\;\;\;\;\;\;\;
\xi_{j-1} \,=\,
\frac{p_{j-1} \cdot p^\prime_{j-1}}{E_{j-1} E^\prime_{j-1}} \,\simeq \,
1 - \cos \theta_{(j\mbox{-}1), (j\mbox{-}1)^\prime}
\;\;\;\;\;\;\;\;\;(m \ge j \ge 1)
\;.
\label{tildek}
\end{equation}
so that the  time-like cascade can be described by
a $\tilde{k}^2$-ordered (rather than $k^2$-ordered) evolution,
which corresponds to an angular ordering with decreasing emission angles
$\theta_{j, j^\prime} > \theta_{(j\mbox{-}1), (j\mbox{-}1)^\prime}$.

Proceeding analogously to the space-like case (c.f. (\ref{PS})),
the probability $d {\cal P}_{m,\,m-1}^{(T)}$ for the
first branching after the $\gamma q$ vertex,
$k_m \rightarrow k_{m-1} k_{m-1}^\prime$
with $k^2_{m-1}, k^{\prime \,2}_{m-1}$,
is given by the space-time extension \cite{ms39,msrep}
of the usual DGLAP probability distribution \cite{DGLAP},
\begin{equation}
d {\cal P}_{m,\, m-1}^{(T)} (z,\tilde{k}^2;\,\Delta t) \,=\,
\frac{d \tilde{k}^2}{\tilde{k}^2}\, dz
\;
\frac{\alpha_s( \kappa^2 )}{2 \pi} \,
\gamma_{m \rightarrow (m\mbox{-}1),(m\mbox{-}1)^\prime} (z)
\;\,
{\cal T}^{(T)} (\Delta t)\,
\; ,
\label{PT}
\end{equation}
where
${\cal T}^{(T)} (\Delta t)$ is the probabilty that parton $m$ with
virtuality
$k_m^2$ and  corresponding proper lifetime $\tau_m \propto
1/\sqrt{k_m^2}$
 decays within a time interval $\Delta t$,
\begin{equation}
{\cal T}^{(T)} (\Delta t)
\;=\; 1\;  - \;\exp \left( - \frac{\Delta t}{t_m(k)}\right)
\label{deltt}
\;.
\end{equation}
The actual lifetime of the decaying parton $m$ in the $ep$ $cms$
is then $t_m(k) = \gamma/\tau_m(k)$, where
$t_q(k) \approx 3 E/(2 \alpha_s k^2)$ for quarks and
$t_g(k) \approx  E/(2 \alpha_s k^2)$ for gluons \cite{ms3}.
As before,
$F_j$ denotes the local density of parton species $j=m,m-1$, and
$\alpha_s/(2\pi \xi) \gamma(z)$ is the DGLAP branching kernel
with energy distribution $\gamma(z)$.
The probability (\ref{PT}) is formulated in terms of the  energy
fractions
carried by the daughter partons,
\begin{equation}
z \;=\; \frac{E_{m-1}}{E_m}
\;\;\;,\;\;\;\;\;\;\;\;
1 - z \; = \; \frac{E_{m-1}^\prime}{E_m}
\;\;\; ,
\end{equation}
with the virtuality $k_m$ of  the quark $m$  related to $z$ and $\xi$
through
$k_m^2 = k_{m_1}^2 + k_{m-1}^{\prime \, 2} + 2 E_m^2 z ( 1 - z ) \xi$,
and the argument $\kappa^2$ in the running coupling $\alpha_s$ in
(\ref{PT})
is \cite{webber88}
$\kappa^2= 2 z^2 ( 1 - z )^2  E_m^2  \xi  \simeq k_\perp^2$.

The branching probability (\ref{PT}) determines the distribution
of emitted partons in both coordinate and momentum space, because
the knowledge of four-momentum and lifetime (or $\Delta t$ between
successive branchings)  give the spatial positions of the
partons, if they are assumed to propagate on straight paths between
the vertices.
The probability that parton $m$ does {\it not} branch between
$\tilde{k}^2$ and
a minimum value $\tilde{k}^2_0 \equiv \mu_0^2$
is given by the exponentiation of (\ref{PT}),
yielding the {\it Sudakov form-factor for time-like branchings}:
\begin{equation}
T_m (\tilde{k}^2,\tilde{k}_0^2; \, \Delta t)
\;=\;
\exp
\left\{
\,-\, \int_{\tilde{k}_0^2}^{\tilde{k^2}}
\, \sum_a
\, \int_{z_{-}(\tilde{k}^\prime)}^{z_{+}(\tilde{k}^\prime)}
\;\,
d {\cal P}_{m,\,m-1}^{(T)} (z, \tilde{k}^{\prime\,2};\,\Delta t)
\right\}
\;,
\label{tlff2}
\end{equation}
which is summed over the species $a = g , q, \bar q$
of parton $m-1$.
The integration limits $\tilde{k}_0^2$ and $z_\pm$
are determined by the requirement that the branching must terminate
when the partons enter the non-perturbative regime and begin
to hadronize.
As we discuss later, this condition can be parametrized by
the confinement length scale $L_c = O(1 \,fm)$  with
$\tilde{k_0}^2  \, \lower3pt\hbox{$\buildrel >\over\sim$}\, L_c^{-2}
\equiv \mu_0^2$, and
$
z_{+} ( \tilde{k}_m ) =
1 -z_{-} ( \tilde{k}_m ) =
\mu_0/\sqrt{4 \tilde{k}_m^2}
$,
so that for
$z_{+} ( \tilde{k}_0^2)=
z_{-} ( \tilde{k}_0^2) = 1/2$
the phase space for the branching vanishes.

The Sudakov form factor (\ref{tlff2}) determines the four-momenta and
positions of the partons of a particular emission vertex as we sketched
above for the first branching, but subsequent branchings are described
completely analogously by replacing $t_m\rightarrow t_{m-1}$,
$x_m \rightarrow x_{m-1}$, $k_m^2 \rightarrow k_{m-1}^2$, etc..
Hence $T(\tilde{k}^2,\tilde{k}_0^2; \, \Delta t)$
generates the time-like cascade
as sequential branchings starting from $t=0$ at the
hard vertex forward in time, until the partons eventually hadronize
as discussed below.
\medskip

\noindent {\bf 3.6  Cluster formation and hadronization}
\smallskip

Both the cluster formation from the collection
of quarks and gluons at the end of the perturbative phase
and the subsequent cluster decay into final hadrons
consist of two components:
\begin{description}
\item[{\bf (i)}]
The recombination of the {\it secondary time-like partons},
their conversion into colourless
{\it parton clusters} and the subsequent
decay into secondary hadrons.
\item[{\bf (ii)}]
The recombination of the {\it primary space-like partons} that remained
spectators throughout the collision development into {\it beam clusters}
and the fragmentation of these clusters.
\end{description}

The important assumption here is that the process of hadron formation
depends only on the local space-, time-, and colour-structure of the
parton system,
so that the hadronization mechanism can be modelled as the formation of
colour-singlet clusters of partons as independent entities
(pre-hadrons),
which subsequently decay into hadrons.
This concept is reminiscent of the `pre-confinement' property
\cite{preconf} of parton evolution,
which is the tendency of the produced partons
to arrange themselves in colour-singlet clusters with limited
extension in both position and momentum space, so that it is
suggestive to suppose that these clusters are the basic units out of
which hadrons form.
\smallskip

\noindent {\bf 3.6.1 Cluster formation}

{\bf (i)  Parton clusters:}
Parton clusters are formed from secondary partons, i.e. those
that have been produced by the hard interaction and the
parton shower development.
The coalescence of these secondary  partons to colour-neutral clusters
has been discussed in detail in Refs. \cite{ms37,ms40}, so that we
confine ourselves here to the essential points.
Throughout the dynamically-evolving
parton shower development, we consider every parton and its nearest
spatial neighbour as a potential candidate for
a 2-parton cluster, which, if colour neutral, plays the role
of a `pre-confined' excitation in the process of hadronization.
Within each single time step, the probability for parton-cluster
conversion is determined for each nearest-neighbor pair by
the requirement that the total colour charge of the two partons must
give a composite colour-singlet state, and the condition that
their relative spatial distance $L$ exceeds the critical
confinement length scale $L_c$.
We define $L$ as
the Lorentz-invariant distance $L_{ij}$ between parton $i$
and its nearest neighbor $j$:
\begin{equation}
L(r_i,r_j)\;=\; L_{ij} \;\equiv \;
\mbox{min} (\Delta_{i 1}, \ldots , \Delta_{i j}, \ldots , \Delta_{i n})
\;,
\label{L}
\end{equation}
where $\Delta_{ij} \equiv \sqrt{ r_{ij}^\mu  r_{ij, \mu} }$,
$r_{ij} = r_i - r_j$,
and the probability for the coalescence of the two partons $i$, $j$ to
form
a cluster is modelled by a distribution of the form
\begin{equation}
\Pi_{ij\rightarrow c}\;\propto \; \left(\frac{}{}
1\,-\, \exp\left(-\Delta F\;L_{ij}\right)
\right)
\;\,\simeq \;\,
1\;-\;\exp\left(\frac{L_0-L_{ij}}{L_c-L_{ij}} \right)
\;\;\;\;\;\mbox{if $L_0 \;<\;L_{ij}\;\le\;L_c$}
\;,
\label{Pi2}
\end{equation}
if $L_0 <L_{ij}\le L_c$, and
0 (1) if $L_{ij} < L_0$ ($L_{ij} > L_c$).
Here $\Delta F$ is the local change in the free energy
of the system that is associated with the
conversion of the partons to clusters,
and the second expression on the right side is our parametrization
in terms of $L_0 = 0.6$ $fm$ and $L_c = 0.8$ $fm$
that define the transition regime.
As we studied in Ref. \cite{ms40}, the aforementioned colour constraint,
that only colourless 2-parton configurations may produce a cluster,
can be incorporated by allowing
coalescence for any pair of colour charges, as determined by the
space-time separation $L_{ij}$ and the probability (\ref{Pi2}),
however, accompanied by the additional emission
of a gluon or quark that carries away any unbalanced net colour
in the case that the two coalescing partons are not in a colourless
configuration.

{\bf (ii)  Beam clusters:}
The remaining fraction of the longitudinal momentum and energy that has
not
been redirected and harnessed by the interaction with the photon is
carried
by the primary partons of the initial proton, which
remained spectators throughout. In our approach
these partons maintain their originally assigned momenta
and their space-like virtualities. Representing the  beam remnant, they
may be pictured as the coherent relics of  the original
proton wavefunction.
Therefore the primary virtual partons must be treated differently
than the secondary partons which are real excitations that contribute
incoherently to the hadron yield.
In the $ep$ $cms$  the primary partons are
grouped together to form a  massive beam cluster with its four-momentum
given by the sum of the parton momenta and its position given by
the 3-vector mean of the  partons' positions.

\noindent {\bf 3.6.2  Hadronization of clusters}

{\bf (i) Parton clusters:}

For  the decay of each parton cluster into final-state hadrons,
we employ the scheme presented in Refs. \cite{ms37,ms40,webber84}:
If a cluster is too light to
decay into a pair of hadrons, it is taken to represent
the lightest single meson that corresponds to its
partonic constituents. Otherwise, the cluster
decays isotropically in its rest frame into
a pair of hadrons, either mesons or baryons, whose combined
quantum numbers correspond to its partonic constituents.
The corresponding  decay probability is chosen to be
\begin{equation}
\Pi_{c\rightarrow h}\;=\;
\;{\cal T}_c(E_c,m_c^2) \;
\;\,{\cal N}
\int_{m_h}^{m_c}
\frac{dm}{m^3}\;\exp\left(-\frac{m}{m_0}\right)
\;,
\label{pi3}
\end{equation}
where ${\cal N}$ is a normalization factor,
and the integrand is a Hagedorn spectrum \cite{hagedorn} that
parametrizes quite well
the density of accessible hadronic states below $m_c$ which are
listed in the particle data tables, and $m_0 = m_{\pi}$.
In analogy to (\ref{deltt}), ${\cal T}_c$ is a
life-time factor giving the probability
that a cluster of mass $m_c^2$ decays  within
a time interval $\Delta t$ in the global frame, here the $ep$ $cms$,
\begin{equation}
{\cal T}_c(E_c,m_c^2)\;=\; 1\;-\;
\exp\left( - \frac{\Delta t}{t_c(E_c,m_c^2)}\right)
\;,
\label{lft2}
\end{equation}
with the Lorentz-boosted life time
$t_c= \gamma_c \tau_c\simeq E_c/m_c^2$.
In this scheme, a particular cluster
decay mode is obtained from (\ref{pi3})
by summing over all possible decay channels,
weighted with the appropriate spin, flavour, and
phase-space factors, and then choosing the actual decay mode
acording to the relative probabilities of the channels.

\smallskip

{\bf (ii) Beam clusters:}

The fragmentation of the beam cluster containing the spectator partons
mimics in our model what is commonly termed the `soft underlying event',
namely, the emergence of those final-state hadrons that are
associated with the non-perturbative physics which underlies the
perturbatively-accessible dynamics of the hard interaction
with parton shower fragmentation.

In the spirit of Ref. \cite{webber88}, we employ a (suitably modified
for our purposes) version of the  soft hadron production model
of the UA5 collaboration \cite{UA5}, which is based on a parametrization
of the CERN $p\bar p$ collider data for minimum-bias hadronic
collisions.
The parameters involved in this model are set to give a good
agreement with those data.

We view  soft hadron production  as a universal mechanism
\cite{Mdiff} that is
common to all high-energy collisions that involve beam hadrons in the
initial state,
and that depends essentially on the total energy-momentum of
the fragmenting final-state beam remnant.
Accordingly, we assume that the fragmentation of the final-state
beam cluster depends solely on its invariant mass $M$, and that it
produces a charged-
particle multiplicity with a binomial distribution \cite{UA5},
\begin{equation}
P(n) \;=\; \frac{\Gamma(n+k)}{n! \Gamma(k)} \;
\frac{ (\overline{n}/k)^n}{(1 + \overline{n}/k)^{n+k}}
\label{ndist}
\;,
\end{equation}
where the mean charged multiplicity
$\overline{n}\equiv\overline{n}(M^2)$
and the parameter $k\equiv k(M^2)$
depends on the invariant cluster mass
\footnote{
Notice that in our model $M$ fluctuates statistically, as a result
of fluctuations of the initial-state parton configuration in
the proton and the variation of the hard scattering variables $x$ and
$Q^2$.
Hence the distribution (\ref{ndist}) and the mean multiplicity
(\ref{nk})
vary from event to event. This is in contrast to the original UA5 model,
in which the fixed beam energy $\sqrt{s}/2$ controls the energy
dependence of
soft hadron production.
}
according to the
following particle data parametrization \cite{UA5},
\begin{equation}
\overline{n}(M^2) \;=\;  10.68 \; (M^2)^{0.115} \;-\; 9.5
\;\;\;\;\;\;\;\;\;
k(M^2) \;=\;  0.029 \; \ln(M^2) \;-\; 0.064
\label{nk}
\;.
\end{equation}
Adopting the scheme of Marchesini and Webber \cite{webber88}, the
fragmentation of a beam cluster of mass $M$ proceeds then as
follows:
First, a particle multiplicity $n$ is chosen from (\ref{ndist}), and the
actual charged particle multiplicity is taken to be $n$ plus the
modulus of the beam cluster charge.
Next, the beam cluster is split into
sub-clusters $(q_1 \bar{q}_2), (q_2 \bar{q}_3), \ldots$ ($q_i = u, d$),
which are
subsequently hadronized in the beam cluster rest frame,
in the same way as the parton clusters described in the preceding
subsection.
To determine the sub-cluster momenta, we assume  a mass distribution
\begin{equation}
P(M) \;=\; c \; (M-1) \; \exp \left[ -a (M-1) \right]
\label{mdist}
\;,
\end{equation}
with $c$ a normalization constant and $a = 2$ GeV$^{-1}$, resulting
in average value of $\langle M \rangle \approx 1.5$ GeV.
The transverse momenta are taken from the distibution
\begin{equation}
P(p_\perp) \;=\; c^\prime \; p_\perp \;
\exp \left[ -b \sqrt{p_\perp^2 +M^2}\right]
\label{pTdist}
\;,
\end{equation}
with normalization $c^\prime$ and slope parameter
$b = 3$ GeV$^{-1}$, and
the rapidities $y$ are drawn from a simple flat distribution
$P(y) \propto const.$ with an extent of
0.6 units  and Gaussian tails with 1 unit standard deviation at
the ends.
Finally, all  hadronization products of the sub-clusters are
boosted from the rest frame of the original beam cluster back into the
global frame, i.e. the $ep$ $cms$.
\bigskip
\bigskip

\noindent {\bf 4. MODEL RESULTS FOR NON-DIFFRACTIVE DIS AT HERA}
\bigskip

\noindent {\bf 4.1 Characteristic evolution of small-{\boldmath{$x$}}
versus large-{\boldmath{$x$}}  scattering events}
\medskip

The kinematics of DIS has very different consequences in
the small-$x$ and large-$x$ regime, as we shall discuss now within our
model.
Specifically, we distinguish
here and in the following the two distinct regimes
\begin{equation}
\mbox{`small' $x$:} \;\;1.7 \cdot 10^{-4} \;\le \; x \;\le \;2.3 \cdot
10^{-3}
\;\;\;\;\;\;\;\;\;\;\;
\mbox{`large' $x$:} \;\; x \;>\; 5 \cdot 10^{-3}
\;,
\label{xrange}
\end{equation}
where the small-$x$ regime is the
typical range probed at HERA, and part of the large-$x$ range
($x  \, \lower3pt\hbox{$\buildrel >\over\sim$}\, 5\cdot 10^{-2}$)
corresponds to previous fixed-target experiments (c.f. Fig. 3).
Table 1 provides the corresponding mass $W$ of the
hadronic system (equal to the total hadronic energy in the $\gamma p$
$cms$),
where the small-$x$ regime is the HERA range which we primarily
focus on in the following.

Figure 6 illustrates vividly the differences between the small-$x$
(left panel) and large-$x$ (right panel)
kinematics for  typical HERA values of $Q^2=8/14/28$ GeV$^2$.
The top plots show the associated
probability distributions for the occurrence of a particular
mass $W$ of the hadronic system produced by the hard interaction
in the specified $Q^2$- and $x$-range.
As could be expected, the differences between the two
$x$-domains are striking: not only the shape, but also the mean
values of the distributions are very distinct (note the different scales
of the
$W$ axis). The most probable $W$ values lie between 100-200 GeV (small
$x$)
and 10-20 GeV (large $x$).
The plots in the middle show the $W$ dependence of the total hadron
multiplicity
calculated within our model. The shape of the curves is
again rather different, which is
a direct consequence of the probability distribution in $W$, and the
phase space available for a given $W$.  On the other hand, the
total numbers of produced hadrons ($N_h \simeq 35 - 40$) in
events around the most  probable $W$ are very similar.
The bottom plots show the corresponding mean values of
the $x_F = 2 p_z/W$, with $p_z$
in the $ep$ $cms$ along the beam axis in the opposite direction
to  the incoming proton, i.e., the
fractional longitudinal momentum carried by the final-state hadrons
which emerge from the parton shower and fragmentation of the struck
quark jet.
Again the $W$ dependences of $\langle x_F \rangle$ are very different
in the small-$x$ and large-$x$ regimes, with the typical $x_F$ ranges
$0.03\, \lower3pt\hbox{$\buildrel <\over\sim$}\,  \langle x_F\rangle
\, \lower3pt\hbox{$\buildrel <\over\sim$}\,  \le 0.13$ and
$\langle x_F \rangle \, \lower3pt\hbox{$\buildrel >\over\sim$}\, 0.5$,
respectively.
\smallskip

The distinct characters of small-$x$ and large-$x$ DIS events in the
kinematic regimes discussed above are accompanied by
different space-time evolution patterns of the
particles in position and momentum space.
As explained in Sec. 3,  our approach allows us to follow the time rate
of change
of the particle densities and associated spectra.
Fig. 7 exhibits the time evolution of the rapidity ($y$) distribution,
and
the particle distributions in longitudinal ($z$) and transverse
($r_\perp$)
direction
with respect to the $ep$ $cms$, our chosen  global frame of description.
The left (right) panel corresponds to small-$x$ (large-$x$) events with
fixed $Q^2=28$ GeV$^2$.
In each plot the three curves correspond to times 0.4/12/20 $fm/c$
after the photon-quark scattering, and each curve includes {\it all}
particles (partons and formed hadrons), which are
actively present in the  mixed particle system at the
specified times.
Comparing the time development of the spectra
$dN/dy$ spectra for  small-$x$ and large-$x$
events, one observes that in the former event type most particles are
produced at central rapidities $\vert y\vert \le 1$,
with a shift toward the proton side (negative $y$), while in the latter
event class this region is least populated.
Related to that, the $dN/dz$ distributions show a much larger
particle production along the beam axis around $z = 0$ $fm$ for
small-$x$ than for large-$x$ events.
>From the $1/N dN/dr_\perp^2$ distribution one can see that
for small-$x$ scattering
the diffusion of the expanding particle system in the transverse
direction
is faster than in the large-$x$ events, an effect that
arises from the transverse pressure of the larger
number of produced particles in the central region.
\bigskip

\noindent {\bf 4.2 Inclusive hadron spectra
in {\boldmath{$x_F$}}, {\boldmath{$p_T^\ast$}},
and  the {\boldmath{$\langle p_T^{\ast\;2} \rangle $}} dependence}
\medskip

The study of
particle multiplicities and momentum distributions of
the hadronic final state at HERA provides sensitive
information about both the QCD processes at the parton level
and the properties of hadron formation. An excellent recent review
can be found in Ref. \cite{pavel}.
One of the attractive features
of the HERA experiments is the production of a large-mass
hadronic final state with $W \simeq $ 100-250 GeV (an order of
magnitude larger than in previous fixed-target experiments).
The conjecture is therefore that the influence of the non-perturbative
QCD effects is less important and that - in the spirit of
`local parton-hadron duality' \cite{LPHD} - the observed hadronic final
state reflects more the dynamics of the partonic processes.

Particularly sensitive measures of the parton level dynamics are the
$x_F$ and $p_\perp^\ast$ distributions, as well as the
$\langle p_\perp^{\ast\;2} \rangle$ of produced charged hadrons,
as measured in the $\gamma p$ $cms$ (\ref{hcmframe}),
where the Feynman variable $x_F = 2 p^\ast_\parallel/W$ and
$p^\ast_\perp$ characterize the momentum
components of hadrons parallel and transverse to the
photon direction.
At large values of $W$ (small values of Bjorken $x$),
these observables are sensitive to hard multigluon radiation.
This feature is evident in Fig. 8, where we plot
our model results for the
$x_F$ and $p^\ast_\perp$ spectra and the dependence of
$\langle p_\perp^{\ast\;2} \rangle$,
for three  typical HERA values $Q^2=8/14/28$ GeV$^2$ and the small-$x$
regime
defined by (\ref{xrange}).
Also shown are the corresponding measured distributions
for $Q^2 = 28$ GeV$^2$ from ZEUS \cite{zeus1}, with which the
calculated dashed-dotted curves ($Q^2 = 28$ GeV$^2$) agree
reasonably well.
All three distributions have a  specific form
due to QCD gluon emission on the parton level which
cannot be explained by the naive `quark parton model' (QPM)
which accounts only for the lowest-order photon-quark scattering
and omits all higher-order QCD radiation. For
comparison, the QPM results are plotted as thin curves.

The $x_F$ distributions (top) show steep exponential decreases
above $x_F \approx 0.05 - 0.1$ and an enhanced particle yield below
that value. Note that the QPM, i.e., the leading-order
Born scattering alone,  gives a slightly shallower decrease.
The effect of the higher-order radiative processes is however
very prominent in the $p_\perp^\ast$ spectra (middle) integrated
over $x_F \ge 0.05$, which
show a power-law dependence due to multi-gluon emission and
a significant contribution of hard gluons with transverse momenta
$\, \lower3pt\hbox{$\buildrel >\over\sim$}\, 3$ GeV (large, in view of
$\sqrt{Q^2} =$ 3-5 GeV).
This result is in vivid contrast with the corresponding QPM result,
which hardly gives any transverse momenta
$\, \lower3pt\hbox{$\buildrel >\over\sim$}\, 1$ GeV.
The mean square of $p_\perp^\ast$, $\langle p_\perp^{\ast\;2}\rangle$
(bottom) is particularly sensitive to the tail of the distribution
and exhibits half of a `seagull' shape for positive values
of $x_F$. The rise of  $\langle p_\perp^{\ast\;2}\rangle$
with increasing $x_F$ is due to the leading hadron effect,
which can be understood as follows:
If $z_h$ denotes the fraction of momentum the initially-struck
quark transferrred to a hadron,
\begin{equation}
\langle p_\perp^{\ast\;2}\rangle  \;=\;
z_h \; \left(\frac{}{}
\langle p_\perp^{\ast\;2}\rangle _{prim} +
\langle p_\perp^{\ast\;2}\rangle _{sec}
\right) \;\,+\;\,
\langle p_\perp^{\ast\;2}\rangle _{frag}
\label{pttot}
\;,
\end{equation}
where
$p_{\perp\;prim}^\ast$ is the primordial transverse momentum
of the quark to which the photon couples (c.f., eq. (\ref{xsec2})),
$p_{\perp\,sec}$ is the secondary contribution from
QCD radiation, and
$p_{\perp\,frag}$ denotes the additional transverse momentum
produced in the fragmentation and hadron formation process,
which is in the average about 0.45 GeV and almost independent of $W$ and
$Q^2$.
>From (\ref{pttot}) one sees that the leading hadrons at
large $x_F$, which carry higher fractional momentum $z_h$,
also carry a higher fraction of the actual parton transverse momentum
(primary plus secondary). This explains qualitatively
the `seagull' shape and the rise of the
$\langle p_\perp^{\ast\;2}\rangle$ of hadrons as a function of $x_F$.
Furthermore, one observes that, the more
gluons with relatively large transverse momentum are radiated,
the larger the contribution  to $\langle
p_\perp^{\ast\;2}\rangle_{sec}$,
and hence the stronger is the effect.
In the QPM with no QCD radiation at all,  the total
$\langle p_\perp^{\ast\;2}\rangle$ is therefore a factor of 5-10 smaller
and has only a weak $x_F$ dependence.
\smallskip

In Fig. 9
the $W$  and $Q^2$ dependences of the mean squared $p_\perp^\ast$ of
hadrons is shown
for two intervals $0.1 < x_F < 0.2$ and $0.2 < x_F < 0.4$,
and compared with data obtained by the ZEUS collaboration.
The agreement with the data is fairly good for the
$Q^2$-dependence, whereas it is less clear for the dependence on $W$.
The $\langle p_\perp^{\ast\;2}\rangle$ depends
strongly on both $W$ (for fixed $Q^2 = 28$ GeV$^2$) and
$Q^2$ (for fixed $W = 120$ GeV).
It is worth noting that in previous fixed-target experiments at lower
energies the $\langle p_\perp^{\ast\;2}\rangle$ is generally
much smaller, depending only weakly
 on $W$ (however at much smaller $W$ values), and
essentially flat in $Q^2$.
\smallskip

We remark that the new class of diffractive events with a large rapidity
gap
is measured \cite{zeus1} to have  very different  $p_\perp^\ast$
spectrum and
$\langle p_\perp^{\ast\;2}\rangle$ from the non-diffractive events that
we have just discussed. The particle distribution in $p_\perp^\ast$
of diffractive events falls much steeper and resembles closely
the QPM curve in the middle plot of Fig. 8.
The mean values $\langle p_\perp^{\ast\;2}\rangle$ are smaller by a
factor
of 2-5, and lie just slightly above the QPM curve in the bottom plot of
Fig. 8.
This indicates that diffractive events resemble in a way
DIS events with very little QCD radiation, consistent with the
common interpretation that the photon couples in these
events to the proton via a colourless intermediate state
which does not fragment by multiparton emision.
\bigskip

\noindent {\bf 4.3 Transverse energy flow}
\medskip

Whereas the inclusive $x_F$  and $p_\perp$ distributions of
produced hadrons are sensitive to the multi-jet structure
due to hard-gluon radiation as discussed above,
the analysis of
inclusive hadron distributions in terms of the global
energy flow extends to classes of events which cannot be
identified unambigously as $n$-jet events.
According to the idea of `local parton-hadron duality' \cite{LPHD},
the pattern of overall distribution of energy among the
partons in an event determines the energy flow observed
in the hadronic spectrum.
The energy flow $dE_\perp^\ast/d\eta^\ast$, and similarly the particle
flow
$dN_h/d\eta^\ast$, are commonly studied as a function of pseudorapidity
$\eta^\ast$ in the hadronic centre-of-mass system, i.e., the $\gamma p$
$cms$,
where
\begin{equation}
\eta^\ast \;=\; - \ln \left(\tan \frac{\theta^\ast}{2}\right)
\;= \;
\ln \left(\frac{E^\ast+ p^\ast_\parallel}{p_\perp^\ast}\right)
\;\;\;\;\;\;\;\;\;\;
E_\perp^\ast\;=\; \sqrt{E^{\ast\;2} + p_\perp^{\ast\;2}}
\label{eta}
\end{equation}
with $\theta^\ast, p^\ast_\parallel, p_\perp^\ast$ defined
with respect to the photon direction.

In Fig. 10 we show  model results
for the distribution of hadrons in $\eta^\ast$, $E_\perp^\ast$ and
as well as the $E_\perp^\ast$ flow.
As before, we choose $Q^2=8/14/28$ GeV$^2$ for the
small-$x$ regime $2.3 \cdot 10^{-4} \le x \le 1.7 \cdot 10^{-3}$
corresponding to $W >$ 60/90/130 GeV (c.f. Table 1).
The plotted distributions reflect the
typical event geography that one expects already from
QPM considerations.
Recall that in the QPM,
with the neglect of higher-order QCD corrections,
the struck quark and the proton remnant system each carry
an energy $W/2$ in the $\gamma p$ $cms$ and move back to back
with rapidities $\pm y_{max}^\ast \propto \pm \ln W/m_\pi$.
The fragmentation of the two receding charges fills the intermediate
pseudorapidity region with hadrons\footnote{For the purposes of
this discussion, we will now neglect the difference between
rapidity $y$ and pseudorapidity $\eta$.}.
The width of the hadron distribution in the final state
is proportional to $\ln W$, while its height is approximately
independent of $W$.
The width of the quark jet and the proton fragmentation
region bounded by $\pm\eta_{max}^\ast$ is typically about 2 units,
corresponding to the $x_F$-range $\vert x_F \vert > 0.05$.
Hence, at high values of $W$ (small $x$), the pseudorapidity range
populated by hadrons can be divided in three regions:
(i) the {\it current jet region} from
$(\eta_{max}^\ast - 2)$ to  $\eta_{max}^\ast$,
(ii) the {\it proton fragmentation region} from
$(-\eta_{max}^\ast + 2$) to  $-\eta_{max}^\ast$,
and (iii) a {\it central plateau region} in between.

The  pseudorapidity distribution $dN_h/d\eta^\ast$  (Fig. 10, top),
as calculated
in our model, shows a distorted version of the naive QPM picture
due to the higher-order QCD radiation effects.
The spectrum is asymmetric and the central plateau is rising
from the proton fragmentation region to the current jet region, rather
than being flat.
Particularly different from the QPM is the behaviour
in the current jet region
$\eta^\ast  \, \lower3pt\hbox{$\buildrel >\over\sim$}\, 3$, which shows
a clear
increase with $Q^2$ of both the height and the width of this part
of the hadron distribution, an effect of the jet broadening due
to gluon radiation off the struck quark.

The  transverse energy distribution $dN_h/dE_\perp^\ast$  (Fig. 10,
middle)
is naturally similar to the $p_\perp^\ast$
spectrum discussed before (c.f. Fig. 8).
It again exhibits a power-law behaviour that is characteristic for
gluon emission with significant $p_\perp^\ast$, a feature which becomes
more
prominent with increasing $Q^2$, because of the enlarged phase-space and
extended duration of parton shower activity  before hadronization.

The hadronic energy flow
$dE_\perp^\ast/d\eta^\ast$  (Fig. 10, bottom) mirrors the distribution
of
energy and transverse momentum among the final-state particles in a
similar way to the pseudorapidity distribution discussed above.
The characteristic features are:
first, a central plateau with a slight dip and a height almost
independent of $Q^2$, secondly, an increase with $Q^2$ of energy deposit
in the
current jet region around $\eta^\ast \approx 3$, resulting from
radiation of the time-like shower of the quark after the hard
scattering,
and thirdly, a similar though much less significant increase with $Q^2$
of the
activity around $\eta^\ast \approx -5$
in the proton fragmentation region due to radiation from the
space-like shower before
the hard interaction, which is going along the proton direction
and opposite to the time-like radiation.
\smallskip

In Fig. 11 the $x$  (or, equivalently, $W$) dependence of the
$E_\perp^\ast$ flow in the current jet hemisphere
in the $\gamma p$ $cms$
is investigated for fixed $Q^2= 28$ GeV.
We plot $dE^\ast_\perp/d\eta^\ast$ for three different
$x$ ranges with average values
$\langle x \rangle = 3.7\cdot 10^{-3}/1.5\cdot 10^{-3}/7.2\cdot
10^{-4}$,
corresponding to $W \simeq 78/137/197$ GeV.
Also depicted are the data points of the measured distributions from the
ZEUS collaboration \cite{zeus2}.
Although the model slightly underestimates the  data
around $\eta^\ast = 0$, qualitative conclusions that may be
drawn are: first, the height
of the plateau-like region for
$\eta^\ast \, \lower3pt\hbox{$\buildrel <\over\sim$}\, 1.5$
is rather independent of $x$, and secondly,  with decreasing
$\langle x \rangle$ (from top to bottom) the peak around the
current jet moves visibly towards larger rapidities,
with $\eta^\ast_{peak} \simeq 2.4 / 2.8 / 3.2$, respectively,
while the height of the peak appears to be stable.
\bigskip

\noindent {\bf 4.4 Mass  distributions of the observed hadronic
final state}
\medskip

As mentioned in the introduction, a new class of DIS events is observed
at HERA in the experiments by ZEUS \cite{lrg1} and H1 \cite{lrg2},
events
which are characterized by a large rapidity gap (LRG) between the proton
and the rest of the hadronic final state that is measured in the
detector.
The properties of these events indicate a diffractive production
mechanism
via exchange of a coherent colourless object between the photon and the
proton
(c.f. Fig. 1b), accompanied by a suppression of QCD radiative processes,
which are, as we discussed, so prominent in non-diffractive events
(c.f., Fig.
 1a)
with no rapidity gap (NRG).
Because in the experiment both diffractive LRG and non-diffractive NRG
events
 are mixed
(with a relative contribution of $\approx$ 5-10 $\%$ from LRG events),
the determination of the diffractive cross-section
requires the detailed knowledge and subtraction of
the non-diffractive contribution.

A method to separate diffractive and non-diffractive contributions
suggested
by ZEUS \cite{zeus3} uses the mass $M_X$ of the hadronic system $X$ that
is
measured in the detector, where
\begin{equation}
M_X^2\;\equiv \;
\left\{\frac{}{}
\left( \sum_j E_j \right)^2 - \left( \sum_j p_{x_j}\right) - \left(
\sum_j p_{y_j}\right)^2 - \left( \sum_j p_{z_j}\right)^2
\right\}_{detector}
\;\; < \;\;W^2
\label{MXdef}
\;,
\end{equation}
includes {\it all} observed particles except for the outgoing electron.
Because of the finite resolution and geometric acceptance
of the detector, $M_X$ is naturally
smaller than the total invariant mass $W$, with the event fluctuations
giving rise to a distribution in $M_X$.
The remarkable feature of
the distributions in $M_X^2$ and $\ln M_X^2$ is that they exhibit very
different
behaviour for the two event types and are sensitive measures
of the event structures.
In this context, we investigate in the following the $M_X$ spectrum of
purely non-diffractive NRG events and its $Q^2$ and $x$ (or $W$)
dependences,
so to provide an estimate of the non-diffractive contribution underlying
the
diffractive LRG component.

Let us briefly summarize the state of knowledge in order to set the
stage.
As illustrated in Fig. 1b,
in {\it diffractive} scattering the outgoing proton or low-mass
nucleonic system
 remains colourless
and escapes through the forward beam hole, while the system $X$ from the
 dissociation of
the photon is, in general, almost fully contained in the detector.
Diffractive dissociation prefers small $M_X$ values
($ \ln M_X^2 \, \lower3pt\hbox{$\buildrel <\over\sim$}\, 4$)
with an event distribution
of the form \cite{zeus3}
\begin{equation}
\frac{d{\cal N}^{diff}}{d\ln M_X^2} \;=\; a \;
\left(\frac{1}{M_X^2}\right)^n
\label{diff}
\;
\end{equation}
where at high energy one expects $n\approx 0$ \cite{Mdiff},
approximately
 independent of
the total $\gamma p$ $cm$ energy $W$.

On the other hand, in {\it non-diffractive} events the incident proton
is broken
 up
and the remnant of the proton is a coloured object with the struck quark
taking
away the net colour. As we discussed before, this results
in a substantial amount of initial- and final-
state radiation followed by hadron formation between the directions of
the
proton and the current jet (c.f., Fig. 1a).
>From perturbative QCD arguments, as well as simple phase-space
considerations,
one expects \cite{zeus3} that the associated event distributions are
peaked
at large $M_X$ values ($ \ln M_X^2 \approx$ 5-10)
with an exponential fall-off towards smaller masses,
\footnote{
Another salient feature of the $\ln M_X^2$ distribution
is an observed scaling in $\ln M_X^2 - \ln W^2$ \cite{zeus3}, implying
that the
 position of the high-mass
peak in $\ln M_X^2$ grows proportional to $\ln W^2$, and the
slope of the exponential fall-off to small $\ln M_X^2$ values is
approximately
independent of $W$.
}
\begin{equation}
\frac{d{\cal N}^{nondiff}}{d\ln M_X^2} \;=\; c\; \exp \left( b\, \ln
 M_X^2\right)
\label{nondiff}
\;,
\end{equation}
where $c$ is a constant. The slope $b$ is the parameter of interest:
on the parton level it can be shown to be determined by the
QCD Sudakov form factor and thus
the probability for gluon emission. In our
model for parton-hadron conversion, it is therefore
closely related to the probability for cluster formation and hadron
production.

In Fig. 12 we show the non-diffractive event distributions in $M_X$
(top) and
$\ln M_X^2$ (bottom). The plots compare calculations with the fixed
values
$Q^2=$ 8/14/28 GeV$^2$, normalized to the total number of events, and
within
the nominal kinematic acceptance of the ZEUS detector,
$-3.8 \le \eta \le 4.3$.
The distribution $1/{\cal N} d{\cal N}/d M_X$ (${\cal N} \equiv {\cal
 N}^{nondiff}$)
in the top part of the figure exhibits a clear $Q^2$ dependence
in both the position of the peak and the extension of the tail towards
large
 $M_X$ values.
The mean values $\langle M_X \rangle = 78 / 93 / 105$ GeV for $Q^2 = 8 /
14 /
 28$
GeV$^2$.
The properties of the distributions are most evident when studied as a
function
 of
$\ln M_X^2$, as in the bottom part where
$1/{\cal N} d{\cal N}/d \ln M_X^2$ is shown for the same parameters.
In this representation the the mass peak exhibits a steep exponential
fall-off towards smaller $M_X^2$ values. The associated slope
exhibits no significant dependence on
$Q^2$ and comes out as $b = 0.95\pm 0.1$, when
compared with the form (\ref{nondiff}) as indicated by the straight line
in the
 plot.
On the other hand, an experimental data analysis by ZEUS \cite{zeus3}
yields
a steeper slope, namely $b^{exp}= 1.46\pm 0.15$.
The discrepancy between $b^{exp}$ and  $b$ can have
various reasons
associated with experimental effects that we did not attempt to
simulate or correct for, e.g., detector acceptance
or other effects, such as energy loss of particles in the calorimeter.
Such effects may affect the value of $b$. However, the fact that
both  $b$ and $b^{exp}$ come out to be universally constant supports the
the conjecture that the difference between the values of $b$ and
$b^{exp}$ is due to global
effects that are missing in our calculations.

In Fig. 13 we investigate the $W$ dependence of the $\ln M_X^2$
distribution
at $Q^2 = 14$ GeV, plotting our results for three distinct
intervals of the total $\gamma p$ $cm$ energy $W$.
One observes that
the slope is the same in all three $W$ ranges, and hence appears to be
 independent
of $W$ as well as $Q^2$.
The position of the peak, however, is shifted to larger values as $W$ is
 increased,
as in the previous figure when $Q^2$ is increased.
Also shown in Fig. 13 are the corresponding measured distributions
measured by ZEUS \cite{zeus3}, with which our model calculations agree
reasonably
 well
for the large $M_X$ range, $\ln M_X^2 \, \lower3pt\hbox{$\buildrel >
 \over\sim$}\, 4-6$,
but which show an additional component at small values
$\ln M_X^2 \, \lower3pt\hbox{$\buildrel < \over\sim$}\, 4$.
This latter is due to the diffractive (LRG) contribution, which
evidently
has a plateau-like (rather than exponential) shape, in agreement with
the expectation (\ref{diff}).
The comparison between the data points and our model results
exhibits two important conclusions.
First, the diffractive component cannot at all be explained by the
standard QCD parton shower evolution plus hadronzation model of
non-diffractive events: in fact, it is
completely absent therein.
Secondly, the diffractive and non-diffractive contributions
appear to be sharply separated when studied with respect to
the variable $\ln M_X^2$, and allow one to  subtract cleanly from
the measured data sample the non-diffractive
part, as calculated using this or other QCD parton shower models.
\medskip

Finally, we would like to comment on the
difference between the value of $b$ calculated within
our model, as compared to
other parton shower models \cite{ariadne,pythia,lepto},
which use the string fragmentation approach \cite{stringmodel}
to hadron formation from the final-state parton ensemble.
As investigated in Ref. \cite{zeus3}, the latter
give a value of $b \simeq 2 $, i.e. about twice as large as in our
model. We believe that this difference arises from
differences in modelling the parton-hadron transition,
i.e., string fragmentation versus cluster formation and decay
\footnote{
The preceding parton shower stage is essentially the same in
our and other models \cite{mc}, with
our additional space-time information becoming relevant during
cluster formation and hadronization.}, as we now discuss in more detail.

Let us first recall some general features of hadron
distributions and correlations within the Mueller approach \cite{mueller},
where they are related by unitarity to appropriate absorptive
parts of forward multiparticle scattering amplitudes.
In the beam fragmentation region, which is relevant to this
discussion, asymptotic properties of the single- and multiparticle
distributions are controlled by Regge singularities. In particular,
the asymptotic value of the single-particle density is controlled
by the pomeron, with subasymptotic corrections and
finite-range multiparticle
correlations controlled by subleading Regge singularities.
These give in particular local two-particle correlations with
a correlation length $\Delta y = 1/{\Delta \alpha}$, where
$\Delta \alpha$ is the difference between the intercepts at
$t = 0$ of the pomeron and the next subleading trajectories,
commonly believed to be the $\rho$ and degenerate trajectories
with $\Delta \alpha \simeq 1/2$, yielding $\Delta y \simeq 2$.
The $\pi$ trajectory with $\Delta \alpha \simeq 1$ would yield shorter-range
correlations with $\Delta y \simeq 1$. These subleading trajectories
would also yield subasymptotic corrections to the $M_X$ distribution:
$b \simeq 2 \Delta \alpha$, corresponding to $b \simeq 1 (2)$ for
the $\rho (\pi)$ trajectories, with the tail corresponding to
rapidity-gap events corresponding to pomeron exchange with
$b \simeq 0$.

With these points in mind, we now recall aspects of particle
production according to the Lund string fragmentation model
\cite{stringmodel},
as compared to our approach. In the former model, the string
is chopped at an ordered sequence of points along the
rapidity axis, with separations chosen randomly but with a mean
value $\delta y \simeq 1$. The string bits then hadronize
independently, with resonance decays
resulting in a correlation length $\Delta y
\simeq \delta y \simeq 1$. The adjacency in rapidity of
the Lund fragmentation model
clearly results in the minimum possible correlation length
$\Delta y$, and hence effectively to the largest possible
$\Delta \alpha$. Thus it is no surprise to find that
simulations based on this model yield a relatively high
value of $ b \simeq 2$, corresponding to the $\pi$ trajectory in
the Mueller language. On the other hand, in our space-time
approach, the pre-hadronic clusters are formed by adjacent
pairs of partons in {\it position} space, which are not
necessarily the closest in rapidity space. This point is
reflected in Fig.~14, where we see that the separation in
rapidity between partons that combine to form a cluster is
typically $\delta y \simeq 2$, approximately a factor $2$
larger than in the Lund model. We therefore expect in our
model that $\Delta y \simeq 2$, corresponding to $\Delta \alpha
\simeq 1/2$ and $b \simeq 1$ as we found above.

To the extent that the experimental value of $b$ exceeds unity,
it may be that our space-time approach deviates too far from the
adjacency in rapidity of the Lund string fragmentation model,
and the truth may well lie somewhere in between, corresponding
in the Mueller approach perhaps to a combination of the $\rho$
and $\pi$ trajectories. One way to
test this would be to measure experimentally the rapidity
correlation length, and compare it directly with the predictions
of various models. An interesting issue to watch will be whether
$b$ and the effective two-particle correlation length depend
on $Q^2$ or $W^2$. The naive Mueller Regge described above has
been derived for incident hadrons, and may require modification at
large $Q^2$.

\bigskip
\bigskip

\noindent {\bf 5. SUMMARY}
\bigskip

We have presented in this paper the application to DIS at HERA of a
model for the
quantum kinetics of multiparticle production that includes the
space-time
development of the parton shower, cluster formation and hadronization.
Compared
with our previous work, novel features include tracking back to the
initial proton
the development of the space-like parton shower prior to its interaction
with the
virtual photon radiated by the electron. Our procedure tracks in space
and time
the emission and evolution during this development of time-like
secondary
partons, as well as the spectator partons in the proton beam
fragment\footnote{This machinery will be applied in the future to more
complicated
situations including $eA$, $pp$, $pA$ and $AA$ collisions.}. As in our
previous
work, the coalescence of partons to form pre-hadronic clusters is
determined
statistically by a spatial criterion motivated by confinement and a
simple
non-perturbative model for hadronization.

Our space-time approach has enabled us to map the history of the
particle
densities and associated spectra, including the rapidity, longitudinal
and
transverse distributions of particles. These hard results may be
compared with
intuitive pictures of the space-time development of hadronic final
states in DIS.
They will also form the basis for the subsequent extension of our
approach to
shadowing and other interesting effects in $eA$ scattering.

We have also explored in our model inclusive hadron spectra in $x_F$ and
$p_T$,
and the transverse energy flow. Although our model reproduces the
general
features of the observed pattern in energy flow, it shares with other
simulations
the tendency to undershoot the data around $\eta^* = 0$. However, the
discrepancy
is not dramatic, and does not make a strong case for the presence of
imortant physical effects not present in the MLLA approach we use here.

Our model provides distributions of the mass $M_X$
of the observed hadronic final state in events without a large rapidity
gap,
which can be used to estimate the background to the cross-section for
LRG events.
We find a spectrum $\sim \exp (b \ln M^2_X)$ with an exponent $b \simeq
1$, which
is not very different from the observed value $b^{\exp} \simeq 1.5$. A
detailed
comparison with the data requires more understanding of detector effects
and
final-state hadron interactions, which goes beyond the scope of this
paper. The
value of $b$ is sensitive to the rapidity density and other properties
of
pre-hadronic clusters, so the relative success of our model, which has
no
parameters adjusted from its previous applications to $e^+e^-$
annihilation,
gives us some hope that we are capturing important aspects of this
physics.

As already mentioned, the new features of our approach introduced in
this paper,
including the space-time treatment of the initial hadronic state, open
the way to
future applications of our model to $eA$, $pp$, $pA$ and $AA$
collisions, where
the novel features associated with high parton densities will become
more marked.
We aim eventually at a unified space-time description of all these
hadronic
processes.
\bigskip
\bigskip

\noindent {\bf ACKNOWLEDGEMENTS}
\medskip

Many thanks to Mark Strikman for his helpful suggestions
and critical remarks on a preliminary version of the manuscript.
This work was supported in part by the D.O.E under contract no.
DE-AC02-76H00016.

\newpage

{\bf TABLE CAPTIONS}
\bigskip

\noindent {\bf Table 1:}

Examples of the kinematic relations (\ref{sxy}),
(\ref{QW}) between the
Bjorken scaling variable $x$, the
the absolute squared invariant mass of the photon $Q^2$, and
the inelasticity variable $y$, as well as the
total invariant mass $W$ of the hadronic system.

\bigskip
\bigskip
\bigskip

{\bf FIGURE CAPTIONS}
\bigskip

\noindent {\bf Figure 1:}

Schematic diagram of particle production in
a) {\it non-diffractive}, and b) {\it diffractive} DIS events. Here
$W$ is the total invariant mass of the produced hadronic system,
$M_X$ is the mass of the {\it observed} final state in the detector, and
in b) $P^\prime$ represents the outgoing (excited) proton
or low-mass nucleonic resonant state.
\bigskip

\noindent {\bf Figure 2:}
Deep-inelastic $ep$ scattering viewed in a) the HERA
{\it laboratory frame} where $e$ and $p$ collide head on, and b) the
$\gamma p$ $cms$, where $\gamma$ and $p$ collide head on.
\bigskip

\noindent {\bf Figure 3:}

Contour plot in the $x$-$Q^2$ plane of the
mass of the produced hadronic system $W$ and the
inelasticity variable $y$ for $\sqrt{s} = 296$ GeV at HERA.
The approximate region of previous fixed-target experiments is
indicated by the shaded area at
$x \, \lower3pt\hbox{$\buildrel >\over\sim$}\, 10^{-2}$.
(From Ref. \cite{zeus1}.)
\bigskip

\noindent {\bf Figure 4:}

Schematics of the components of our model
for DIS: The
highly Lorentz-contracted  incoming  proton with its initial-state
parton configuration evolves from the remote past $t=t<0$, and
is struck by the photon at $t=0$. This hard interaction picks a quark
out of the proton's parton cloud, thereby triggering
initial-state (space-like) and final-state (time-like)
parton showers. With increasing time $t\rightarrow +\infty$,
the partons evolve by further radiation, whereas
the remnant proton propagates on as a
coherent remainder. In the process of hadronization
the produced partons may coalesce to colourless clusters
if they are nearest neighbours
in space-time, whereas the virtual partons of the proton remnant
combine with a colour-neutralizing parton to form a massive beam
cluster.
Both  `parton clusters' and the `beam cluster'
subsequently convert into primary hadrons that subsequently decay to
low-mass final-state particles.
\bigskip

\noindent {\bf Figure 5:}
Schematic diagram of the probabilistic parton evolution
in the MLLA framework
(solid lines are quarks, curly lines are gluons).
The initial-state quark with
space-like virtuality $-p_0^2 \approx Q_0^2$ evolves
from $t=t_0$ forward in time and toward the hard $\gamma q$ vertex by
successively increasing
its off-shellness up to $-p_n^2 \approx Q^2$, when it is struck by
the photon at $t=0$. The outgoing quark is provided by the momentum
transfer
with a time-like virtuality $k_m^2 \approx Q^2$, and radiates
off its excitation by successive gluon radiation
until it starts to hadronize by coalescence with another parton,
at which point the shower terminates naturally.
\bigskip

\noindent {\bf Figure 6:}

Characteristic differences in global properties
between  `small'-$x$ (left panel) and
`large'-$x$ DIS events (right panel), as defined in (\ref{xrange}).
The two ranges refer to the values of Bjorken $x$ of the
quark struck by the photon with selected $Q^2$. The small-$x$ range
is typical for the kinematics of HERA experiments, whereas the
large-$x$ regime corresponds to the phase-space region
probed by previous fixed-target experiments.
Compared are a) the
probability distributions for the production of a hadronic system
of mass $W$ (top),
b) the corresponding $W$ dependence of the total hadron
multiplicity $N_h$ (middle), and c)
the resulting mean values of  $x_F = 2 p_z/W$, with $p_z$
in the $ep$ $cms$ along the beam axis in the opposite direction
to the incoming proton.
\bigskip

\noindent {\bf Figure 7:}

Characteristic differences in the space-time evolution pattern
of  `small'-$x$ (left panel) and
`large'-$x$ DIS events for DIS events at $Q^2 =28$ GeV
(in correspondence with Fig. 6).
Compared  at 3 different times $t=0.4/12/20$ $fm/c$ are
a) the rapidity distribution $dN/dy$ (top),
b) the particle distribution along the $ep$ beam axis,
$dN/dz$, and
c) the particle distribution perpendicular to the beam axis,
$1/N dN/dr_\perp$.
Note that the distributions include
all particle species (partons and hadrons) present in the system
at the quoted given times.
 However, the $y$ distributions
only count the secondary particles and exclude
the primary partons of the original proton, beacuse their
rapidity is not well defined,
whereas the $z$ and $r_\perp$ spectra include also
those primary partons.
\bigskip

\noindent {\bf Figure 8:}

Model results for differential charged hadron multiplicities
with respect to the $\gamma p$ $cms$
as a function of
$x_F$ and $p_\perp^\ast$ for $x_F > 0.05$, as well as
$\langle p_\perp^{\ast\;2}\rangle$ as a function of
$x_F$.
We compare  results for the small-$x$ regime (\ref{xrange})
at  typical HERA values of $Q^2 =8/14/28$ GeV$^2$.
The data points are measured distributions from
ZEUS \cite{zeus1} for $Q^2 = 28$ GeV$^2$, and the thin solid curves
represent the corresponding expectations of the
`naive' quark parton model (QPM).
\bigskip

\noindent {\bf Figure 9:}

Model results for
the $W$ dependence for fixed $Q^2 =$ 28 GeV$^2$
and the $Q^2$ dependence for fixed $W=$ 120 GeV of the
mean squared transverse momentum of charged hadrons
$\langle p_\perp^{\ast\;2}\rangle$. The
plots refer to the $\gamma p$ $cms$ and separate
the two intervals  $0.1 < x_F < 0.2$ and $0.2 < x_F < 0.4$.
The data points are from  the ZEUS experiment \cite{zeus1}.
\bigskip

\noindent {\bf Figure 10:}

Hadron distributions in the $\gamma p$ $cms$ as
a function of pseudorapidity $\eta^\ast$ and of transverse
energy $E_\perp^\ast$, as well as the $E_\perp^\ast$ flow
versus pseudorapidity.
The model results refer to
$Q^2= 8/14/28$  GeV$^2$ and
$2.3 \cdot 10^{-4} \le x \le 1.7 \cdot 10^{-3}$
corresponding to $W > 60/90/130$ GeV.
\bigskip

\noindent {\bf Figure 11:}

The $x$  ($W$) dependence of the
$E_\perp^\ast$ flow on the current jet side in the $\gamma p$
$cms$ at $Q^2= 28$ GeV and for
$\langle x \rangle = 3.7\cdot 10^{-3}/1.5\cdot 10^{-3}/7.2\cdot
10^{-4}$,
corresponding to $W \simeq 78/137/197$ GeV.
The experimental data are from ZEUS \cite{zeus2}.
\bigskip

\noindent {\bf Figure 12:}

Comparison of normalized event distributions in
$M_X$ (top) and $\ln M_X^2$ (bottom)
at $Q^2=$ 8 / 14 / 28 GeV$^2$, and
for $2.3 \cdot 10^{-4} \le x \le 1.7 \cdot 10^{-3}$.
with $-3.8 \le \eta \le 4.3$.
\bigskip

\noindent {\bf Figure 13:}

Model results for the $W$ dependence of the $\ln M_X^2$ distribution
at $Q^2 = 14$ GeV  for three $W$ intervals, compared with
the corresponding measured distributions from  ZEUS \cite{zeus3}.
\bigskip

\noindent {\bf Figure 14:}
{\it Top:} Population of rapidity $y^\ast$ in the $\gamma p$ $cms$ of
pre-hadronic clusters formed from
coales|ced parton pairs in the current jet region.
{\it Bottom:}
Distribution in relative rapidity $\delta y^\ast = y_1^\ast - y_2^\ast$
of parton pairs making up the clusters.
The thin full line represents a constant rapidity separation
$\delta y^\ast =1$ and serves as reference to the discussion in the text.

\newpage

$ $
\vspace{4cm}

\begin{tabular}{|c|c|ccc|}
\hline
\hline
$\;\;\;\;\;\;\;\;\;\;\;\;\;\;\;\;\;$  &  $\;\;\;\;\; Q^2 \;\;\;\;\;\;$ &
 \multicolumn{2}{c|}{$\;\;\;\;\;\;\;\;\;\;\;$ `small' $x$
 $\;\;\;\;\;\;\;\;\;\;\;\;\;$}
& $\;\;\;\;\;\;\;\;\;$ `large' $x$ $\;\;\;\;\;\;\;\;\;$ \\
\hline
      &(GeV$^2$)&  $2.33 \cdot 10^{-4}$ & $1.72 \cdot 10^{-3}$ &
$5\cdot
 10^{-2}$\\
\hline
\hline
       &   4     &  0.20 &   0.027& 0.0009 \\
       &   8     &  0.39 &   0.053& 0.0018 \\
$y$    &   14    &  0.69 &   0.093& 0.0032\\
       &   28    &  1    &   0.18 & 0.0064\\
       &   54    &  1    &   0.36 & 0.012 \\
       &  110    &  1    &   0.73 & 0.025  \\
\hline
       &   4     &  131  &   48  &  9 \\
$W$    &   8     &  185  &   68  & 12 \\
(GeV)  &  14     &  245  &   90  & 16  \\
       &  28     &  296  &  127  & 23  \\
       &  54     &  296  &  177  & 32  \\
       & 110     &  296  &  253  & 46 \\
\hline
\hline
\end{tabular}
\bigskip
\bigskip
\begin{center}
{\Large {\bf Table 1}}
\end{center}

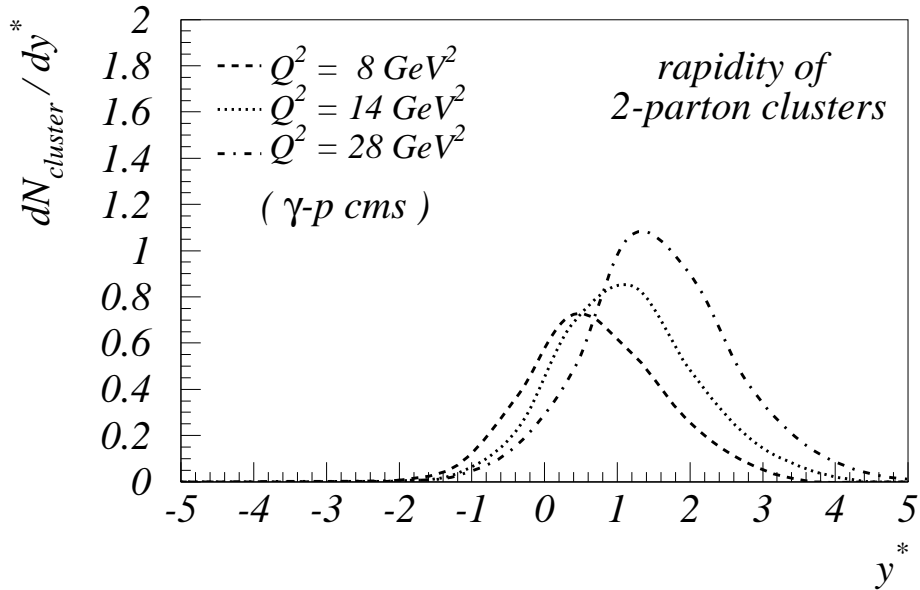
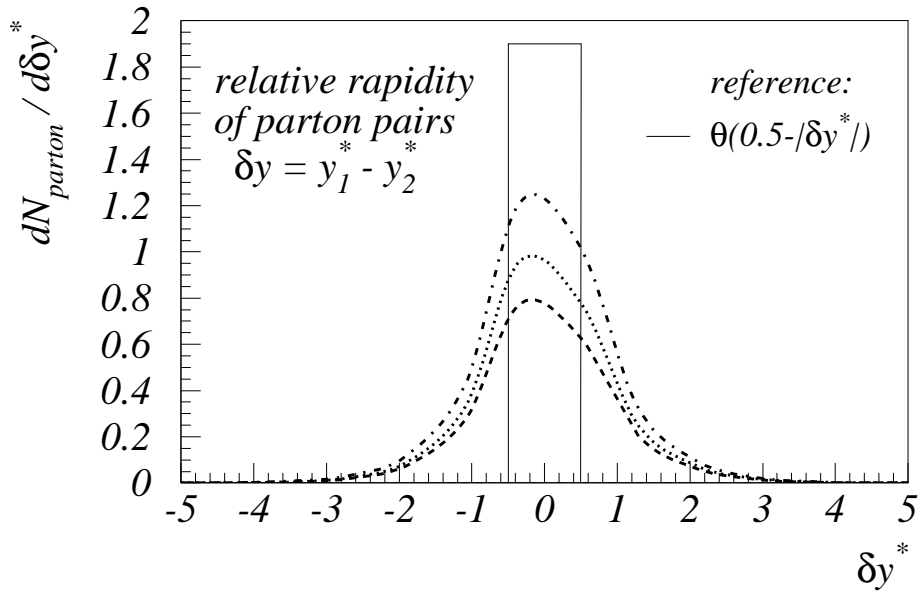
\end{document}